\documentclass[twocolumn]{aastex631}

\usepackage[T1]{fontenc}

\usepackage{natbib}
\usepackage{amsmath}
\usepackage{xcolor}
\usepackage{comment}

\hypersetup{linkcolor=blue,citecolor=blue,filecolor=cyan,urlcolor=blue}

\submitjournal{ApJ}

\shorttitle{Non-thermal radiation from BLRs}
\shortauthors{M\"uller et al.}

\begin{document}

\title{Non-thermal emission from fall-back clouds in the Broad-Line Region of Active Galactic Nuclei}

\correspondingauthor{Ana Laura M\"uller}

\author[0000-0002-8473-695X]{Ana Laura M\"uller}
\affiliation{ELI Beamlines, Institute of Physics, Czech Academy of Sciences, CZ-25241 Doln\'i B\v{r}e\v{z}any, Czech Republic}\email{analaura.muller@eli-beams.eu}

\author[0000-0002-7604-9594]{Mohammad-Hassan Naddaf}
\affil{Center for Theoretical Physics, Polish Academy of Sciences,  Al. Lotnik\'ow 32/46, 02-668 Warsaw, Poland}
\affil{Nicolaus Copernicus Astronomical Center, Polish Academy of Sciences, Bartycka 18, 00-716 Warsaw, Poland}

\author[0000-0001-6450-1187]{Michal Zaja\v{c}ek}
\affil{Department of Theoretical Physics and Astrophysics, Faculty of Science, Masaryk University, Kotl\'a\v{r}sk\'a 2, CZ-611\,37 Brno, Czech Republic}
\affil{Center for Theoretical Physics, Polish Academy of Sciences, Al. Lotnik\'ow 32/46, 02-668 Warsaw, Poland}

\author[0000-0001-5848-4333]{Bo\.zena Czerny}
\affil{Center for Theoretical Physics, Polish Academy of Sciences, Al. Lotnik\'ow 32/46, 02-668 Warsaw, Poland}
 
\author[0000-0001-7605-5786]{Anabella Araudo}
\affil{Laboratoire Univers et Particules de Montpellier (LUPM) Universit{\'e} Montpellier, CNRS/IN2P3, CC72, place Eug{\`e}ne Bataillon,\\ 34095, Montpellier Cedex 5, France}
\affil{ELI Beamlines, Institute of Physics, Czech Academy of Sciences, CZ-25241 Doln\'i B\v{r}e\v{z}any, Czech Republic}
\affil{Astronomical Institute, Czech Academy of Sciences,
Bo\v{c}n\'{\i} II 1401, CZ-141\,00 Prague, Czech Republic}

\author[0000-0002-5760-0459]{Vladim\'ir Karas}
\affil{Astronomical Institute, Czech Academy of Sciences, Bo\v{c}n\'{\i} II 1401, CZ-141\,00 Prague, Czech Republic}

\title{Non-thermal emission from fall-back clouds in the Broad-Line Region of Active Galactic Nuclei}

\begin{abstract}
The spectra of active galactic nuclei exhibit broad-emission lines that presumably originate in the Broad-Line Region (BLR) with gaseous-dusty clouds in a predominantly Keplerian motion around the central black hole. Signatures of both inflow and outflow motion are frequently seen. The dynamical character of BLR is consistent with the scenario that has been branded as the Failed Radiatively Accelerated Dusty Outflow \citep[FRADO;][]{czerny2011}. In this scheme, frequent high-velocity impacts of BLR clouds falling back onto the underlying accretion disk are predicted. The impact velocities depend mainly on the black-hole mass, accretion rate, and metallicity and they range from a few km s$^{-1}$ up to thousands of km s$^{-1}$. Formation of strong shocks due to the collisions can give rise to the production of relativistic particles and associated radiation signatures. In this work, the non-thermal radiation generated in this process is investigated, and the spectral energy distributions for different parameter sets are presented. We find that the non-thermal processes caused by the cloud impacts can lead to emission in the X-ray and the gamma-ray bands, playing the cloud density and metallicity a key role.

\end{abstract}


\section{Introduction}\label{sec:intro}

The spectra of active galactic nuclei (AGN) display distinctive features across a broad range of frequencies, suggesting that these regions contain many different components \citep[see e.g.,][]{antonucci1993,urry1995,2021bhns.confE...1K}. Nevertheless, the inability of resolving such compact and bright regions using the current instruments challenges unveiling the AGN structure in detail. In particular, the characteristic broad-emission lines of type-I AGN \citep{1943ApJ....97...28S,1959ApJ...130...38W,1963Natur.197.1040S} are assumed to be produced very close to the central supermassive black hole (BH), in the so-called broad-line region (BLR). According to the observational data, BLRs are made up of clouds moving primarily in Keplerian orbits \citep[see e.g.,][]{peterson1998,shapovalova2010,grier2013}. Additionally, the measured lines indicate the presence of inflowing and outflowing matter \citep[see e.g.,][]{brotherton1994,done1996,doroshenko2012}, confined in a flattened region \citep{gravity2018}. Although the origin of these line-emitting clouds remains unknown, this particular geometry hints that they are likely related to the accretion disk, at least pertaining to the clouds of the low-ionization line BLR part.

Over the years, several models were developed to explain the formation, dynamics, and emission lines of BLRs \citep[][]{osterbrock1978,blandford1982,begelman1983,murray1995,done1996,hu2008,collin1999,wang2011,czerny2011,baskin2018}. These clouds have also been considered to participate in processes that can produce and absorb high-energy photons \citep{dar1997,araudo2010,delpalacio2019,tavecchio2012,boettcher2016,2021MNRAS.503.3145B}. However, in most of these studies, the geometry of the BLR is assumed to be spherical and the radial distribution of clouds uniform. 

In this work, we aim to study the non-thermal features connected with a realistic flattened BLR description. Particularly, we discuss the BLR properties as expected from the Failed Radiatively Accelerated Dusty Outflow (FRADO) model \citep{czerny2011,czerny2015,czerny2016,czerny2017,naddaf2020, naddaf2021, naddaf2021b}. In this model, the formation of BLR clouds is governed by the radiation pressure of the accretion disk acting on dust grains, analogously as it is inferred in dusty stellar winds, e.g., in AGB stars \citep[see e.g.,][]{Sedlmayr1995}. At distances where the temperature of the accretion disk drops below the dust sublimation temperature, the local radiation pressure is large enough to expel material from the disk. Once the cloud elevates above the disk, the photons coming from the hot inner part of the AGN irradiate the plasma, increasing the temperature, and producing the sublimation of the dust. Without dust, the cloud is not sensitive to the radiation pressure anymore and starts to move by following a ballistic motion, first upwards and finally falling back onto the accretion disk. The production of shocks and non-thermal emission in such cloud impacts was recently proposed and studied by \citet{muller2020}. In their work, the authors found that the emitted radiation can significantly contribute to the X-ray and gamma-ray emission of AGNs.

\begin{figure} 
	\centering
	\includegraphics[width=\columnwidth]{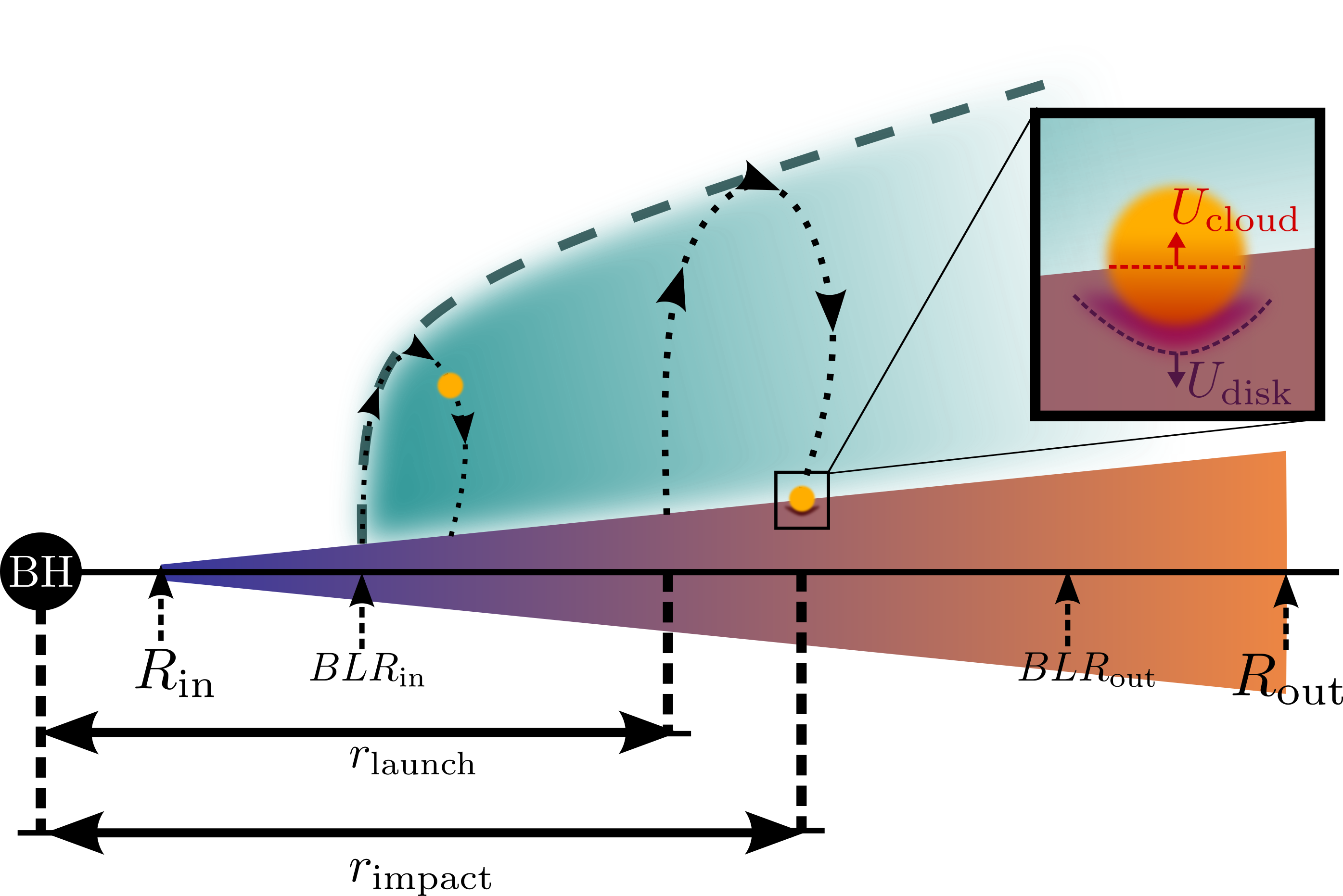}
	\caption{Sketch of the astrophysical situation (drawn not to the scale). In this figure, $R_{\rm in}$ and $R_{\rm out}$ indicate the inner and outer radii of the accretion disk, respectively, whereas $BLR_{\rm in}$ and $BLR_{\rm out}$ are the limits of the BLR. $U_{\rm cloud}$ and $U_{\rm disk}$ are the velocities of the shock waves propagating in the cloud and the disk, respectively.}
	\label{fig:sketch}
\end{figure}

In this paper, we investigate the non-thermal characteristics within the BLR in the scenario investigated by \citet{muller2020} (see Fig. \ref{fig:sketch}), when the cloud dynamics and impacts are described by the 2.5D FRADO model \citep{naddaf2021}. In Section \ref{sec:FRADO}, we refer to the FRADO model and display the parameters explored in our analysis. We examine the feasibility of accelerating particles due to disk-cloud collisions in Section \ref{sec:non-thermal}. Our results are shown in Section \ref{sec:results}. Finally, Section \ref{sec:conclusions} presents a summary and our conclusions. The implications of the assumptions we made, in particular the cloud stability, are discussed in the Appendix~\ref{sec:discussion}.

\section{The BLR dynamics in the FRADO model} \label{sec:FRADO}

The FRADO model describes the dynamics of the BLR clouds in a non-hydrodynamic simple way. A fully analytical form of this approach was first proposed by \citet{czerny2011}, and later developed in more detail \citep{czerny2015,czerny2017}. The description of the full 3D cloud motion requires semi-numerical calculations which were recently presented in \citet{naddaf2020, naddaf2021, naddaf2021b}. The clouds are considered to be launched by the dust-driven outflow. Nevertheless, the rise of their height above the disk increases the exposure to the radiation from the disk central parts, causing the dust sublimation. This effect is taken into account in the model, which reproduces reliably the radius-luminosity relation in the low-ionization part of the BLR, at the basis of the assumed value of the dust sublimation temperature.

We calculate the dynamical properties of the BLR as described in \citet{naddaf2021}. The accretion disk is modeled using a Shakura-Sunyaev disk model with a proper account of the disk vertical structure and internal opacity \citep{rozanska1999,czerny2016} which is important for an accurate determination of the disk thickness as a function of radius. The clouds are ejected from the disk surface and move under the influence of the radiation pressure acting on the dust inside the cloud. The radiation pressure comes mainly from the patch of the disk underlying the cloud, i.e., it is a function of the cloud location and increases with the cloud height. This approach takes into account the shielding of the central source by other clouds and by the inner line-driven wind. 

The innermost radius of the BLR ($BLR_{\rm in}$) is found by requiring the radiation pressure to be strong enough to lift the clouds but the temperature must still be below the sublimation threshold. For given values of BH mass ($M_{\rm BH}$) and accretion rate ($\dot{M}$), this $BLR_{\rm in}$ depends then on the assumed value of the dust sublimation temperature. The computational grid covers the disk densely, with a constant step, linear in radius. The outer edge of the BLR ($BLR_{\rm out}$) is set by the location of the dusty/molecular torus. When the cloud is launched, we follow its trajectory. At each trajectory point, the dust temperature in the cloud is obtained from the dust thermal equilibrium. Once this temperature exceeds the sublimation temperature, the radiation pressure is turned off, and the cloud follows a ballistic trajectory. It is important to note that the gas temperature in the cloud ($\sim 10^4$ K) is generally much higher than the dust temperature, i.e., there is no thermal equilibrium between the dust and the gas due to the large difference in their heating and cooling mechanisms. However, we assume an efficient dust and gas Coulomb coupling, thus the gas hitchhikes with dust.

\subsection{Model Parameters}

We consider two exemplary values of the BH mass: $M_{\rm BH} = 10^6$ and $10^8M_{\odot}$, appropriately covering the most common parameter range observed in AGN. As the FRADO model is sensitive to the Eddington ratio $\dot{m}=\dot{M}/\dot{M}_{\rm Edd}$, we study three different values: 0.01, 0.1 and 1.0, as in \citet{naddaf2021, naddaf2021b}. Since the observational evidence has pointed out that the metallicity $Z$ in AGN is likely super-solar \citep[see e.g.][]{hamann1992, artymowicz1993, ferland1996, hamann1997, dietrich2003, xu2018, sniegowska2020}, we adopt two extreme values: $Z=Z_{\odot}$ and $5 Z_{\odot}$, where $Z_{\odot}$ is the solar metallicity. Specifically, $Z$ enters the model by assuming the dust to gas ratio 0.005 in the basic case as inferred by the classical Mathis-Rumpl-Nordsieck (MNR) dust model \citep{mathis1977}, and 0.025 for the largest metallicity. The assumption $Z > Z_{\odot}$ enhances the radiation pressure acting on the dust, and leads to a more vigorous outflow.

In all the models, we use the wavelength-dependent dust opacity from \cite{rollig2013}, adopting a mixture of graphite and silicate grains with the standard range of grain sizes \citep[see][for details]{naddaf2021}. We assume that the dust sublimation temperature is 1500~K. Cloud launching requires some shielding, and for all the models we fixed the shielding parameter ($\alpha$) describing the fraction of the disk surface seen by the cloud at each point of its trajectory at $\alpha$=3 \citep[see][for more details]{naddaf2021}.

For the purpose of this work, we adopt cloud number densities $n_{\rm c} =10^{11}, 10^{12}\,{\rm cm^{-3}}$ and hydrogen column densities $N_{\rm H} = 10^{23}, 10^{24}\,{\rm cm^{-2}}$, respectively, from which we estimate the characteristic cloud length-scale, $R_{\rm c}\sim N_{\rm H}/n_{\rm c} =  10^{12}\,{\rm cm}$. The characteristic cloud mass then is $M_{\rm c} =  (4/3)\,\pi\, R_{\rm c}^3\, \mu\, m_{\rm p}\, n_{\rm c}\sim 2\times 10^{-9}\,M_{\odot}\sim 4\,M_{\rm Ceres}$, where $M_{\rm Ceres}$ is the Ceres mass.
A summary of the important initial parameters of the model is provided in Tables \ref{tab:cloud_parameters} and \ref{tab:disk_parameters}. In the latter, $r_{{\rm g}_{x}}$ and $\dot{M}_{{E}_{x}}$ are the gravitational radius and the Eddington accretion rate for a BH of $10^{x}$ M$_{\odot}$, respectively.

\subsection{Cloud Fall-Back onto the Accretion Disk}
\label{sect:fall_back}

In the low Eddington accretion regime, all clouds form a failed wind, and their departure from the launching radius becomes relatively small. Consequently, the motion of such clouds under the radiation pressure acting on dust is well approximated by the plane-parallel approach, as argued by \citet{czerny2015}. Since the clouds cover the disk surface uniformly, albeit each of them at a different phase of motion, their impact rate onto the disk can be determined by the duration of the orbital motion. However,  the plane-parallel approach cannot be applied for cases with high Eddington rates. This is because a stream of high velocity clouds forms close to the inner BLR radius for such cases, and those clouds either hit the torus, or all fall down onto the disk very far away from their starting point.

Although the model determines precisely the velocity field of the clouds, it cannot predict directly the local stream of the launched material. To overcome this difficulty, we divide the clouds into two populations: local clouds which form a local failed wind, mostly observed in the outer disk part, and non-local clouds from a stream which can be described using a standard approach like in stellar or disk winds. We define these two categories based on the relative difference between the starting radius of the clouds, $r_{\rm launch}$, and their landing radius, $r_{\rm impact}$. Non-local clouds should satisfy

\begin{equation}
    \frac{|r_{\rm impact} - r_{\rm launch}| }{ r_{\rm launch} } > 0.1.
\end{equation}

\noindent This criterion allows us to distinguish both populations and calculate their impact rate onto the accretion disk. The existence of one or two groups of clouds depends on the parameters adopted for a specific model. For instance, in cases with a low Eddington rate and/or low metallicity, only the local clouds are present.

\subsection{Impact Rate}

In the case of the local clouds, we assume that they cover uniformly the disk surface while performing their up-and-down motion \citep{czerny2015}. Numerically, we consider trajectories with starting points $r_{\rm launch}$ distributed equidistantly along the disk surface. For each trajectory, we compute the time $t_{\rm flight}$ needed to complete the path between the launching and landing points. We thus calculate the impact rate $\dot{N}$ per unit surface of local clouds at the impact distance $r_{\rm impact}$ as
\begin{equation}
\dot N(r_{\rm impact}) =
\frac{1}{t_{\rm flight}\,R_{\rm c}^2 },
\end{equation}

\noindent where $R_{\rm c}$ is the radius of the cloud.

For the non-local clouds, the assumption of the uniform coverage is not appropriate since their launching mechanism is similar to that of an escaping wind. Therefore, we apply the approach developed by \citet{murray1995} for line-driven winds. We assume that the total local radiation flux $F(r)$ at the launching radius $r_{\rm launch}$ is used by the cloud to get an initial velocity equal to the local sound speed $C_{\rm s}$ in the disk atmosphere. Under this condition, the impact rate of non-local clouds can be written as
\begin{equation}
 \dot{N}(r_{\rm impact}) = \frac{F(r_{\rm launch}) }{ c\, C_{\rm s}\, m_p\,n_{\rm c}\, R_{\rm c}^3}\;f(r_{\rm launch},r_{\rm impact}),
 \label{eq:impact}
\end{equation}

\noindent where $f(r_{\rm launch},r_{\rm impact})$ is defined as the ratio of the surfaces of the launching ring to the corresponding impact ring, which can be very different in this case. To adopt  $C_{\rm s}$ instead of escaping velocity of the wind may seem as a serious limitation, but the driving flux increases as the cloud rises higher above the disk and becomes exposed to the strong central radiation accelerating the motion. The non-local clouds, if present, actually fall onto the entire disk surface and could interact with the local population. Nevertheless, our simple model does not account for this possibility. 

The parameters of the cloud collision with the disk are calculated by comparing the ram pressure exerted by the cloud with the local pressure inside the disk. The ram pressure of the cloud is calculated from the formula
\begin{equation}
P_{\rm ram}  = n_c\,m_p\,v_z^2.  
\end{equation}
The velocity $v_z$ is calculated in the local frame co-moving with the disk, i.e., the local Keplerian speed is subtracted. The disk height and the disk vertical structure needed to estimate the depth of the cloud penetration are calculated as in \citet{rozanska1999,czerny2016}, however, without introducing the hot corona or self-gravity to the disk structure \citep{1989ApJ...341..685S,2004CQGra..21R...1K}. All other effects are included, i.e., the proper description of the opacity, including the dust and molecules \citep[see the description in][]{naddaf2021}, convection, and gas and radiation pressure. With the pressure profile inside the disk obtained, $P \equiv P(z)$, we compute the impact depth from the condition
\begin{equation}
    P_{\rm ram} = P(z_{\rm impact}),
    \label{eq:pram}
\end{equation}

\noindent where we introduce the local density and the temperature from the disk structure. Computations are performed at the radius corresponding to the cloud impact position, neglecting the deceleration of the cloud across the disk as well as the change of its distance from the center when passing through the disk interior. The viscosity parameter $\alpha_{\rm accre}$ introduced by \citet{SS1973} is assumed to be 0.02, as estimated at the basis of variability \citep[see][and the references therein]{grzedzielski2017}.

We find that clouds in models adopting the solar metallicity are stopped by the disk, usually relatively close to the disk surface. However, for high metallicity models, the condition given in Eq.~(\ref{eq:pram}) is never met, because the ram pressure of some clouds is higher than the equatorial disk pressure. This means that such clouds should traverse the disk and reappear on the other side. In this case, we use the equatorial values to characterize the disk density and calculate the shock conditions in the next section.

  \begin{table} 
  	\centering
  	\caption{Initial parameter values of a cloud.}
  	\begin{tabular}{l c}
  		\hline\hline
  		Parameter [units] & Value \\ \hline
  		$N_{\rm H}$ column density [cm$^{-2}$] & $10^{23}, 10^{24}$  \\
  		$n_{\rm c}$ number density [cm$^{-3}$] &  $10^{11}, 10^{12}$  \\
  		$R_{\rm c}$ cloud radius [cm] &  $10^{12}$  \\
  		$M_{\rm c}$ cloud mass [M$_{\odot}$] &  $3 \times10^{-9}$ \\
  		$v_{\rm c}$ cloud velocity [km s$^{-1}$] & FRADO \\ \hline
  	\end{tabular}
  	\label{tab:cloud_parameters}
  \end{table}
  
    \begin{table} 
  	\caption{Values of the parameters of the central BH and the associated accretion disk in the model.}
  	\resizebox{\columnwidth}{!}{
  	\begin{tabular}{l c}
  		\hline\hline 
  		Parameter [units] & Value \\ \hline
  		$M_{\rm BH}$ [M$_{\odot}]$ & ${10^{6}}, 10^{8}$ \\
  		${r_{\rm g_{6}}}$ [pc] & ${4.8\times10^{-8}}$ \\
        ${r_{\rm g_{8}}}$ [pc] & $4.8\times10^{-6}$ \\
  		${\dot{M}_{\rm E_{6}}}$ [M$_{\odot}$ s$^{-1}$] & $7\times10^{-10}$ \\
  		${\dot{M}_{\rm E_{8}}}$ [M$_{\odot}$ s$^{-1}$] & $7\times10^{-8}$ \\
        $\dot{m}$ Eddington ratio (${\dot{M}/ \dot{M}_{\rm E}}$) & $0.01, 0.1, 1$ \\
  		$\alpha_{\rm accre}$ viscosity parameter & $0.02$ \\
  		$Z$ metallicity [$Z_{\odot}$] & $1, 5$\\ \hline
  		$r_{\rm impact}$ impact distance [cm] & FRADO \\
  		$\rho_{\rm disk}(r_{\rm impact})$ volumetric density [g cm$^{-3}$] &  FRADO  \\
  		$n_{\rm disk}(r_{\rm impact})$ number density [cm$^{-3}$] &  FRADO  \\  
  		$T_{\rm disk}(r_{\rm impact})$ disk temperature [K] & FRADO \\
  		  $L_{\rm disk}$ disk luminosity [erg s$^{-1}$] & ${\dot{m}~ (\dot{M}_{\rm Edd}\ c^2) / 12 }$ \\
  		  $R_{\rm BLR}$ BLR effective radius [cm] & $(BLR_{\rm out}+BLR_{\rm in})/2$ \\
  		    \hline
  	\end{tabular}}
  	\label{tab:disk_parameters}
  \end{table} 

\subsection{Shock Properties and Non-Thermal emission}\label{sec:non-thermal}

The supersonic collision of a BLR cloud with the accretion disk produces a forward shock in the disk and a reverse shock in the cloud.
The velocities of these shocks can be calculated as \citep[see e.g.,][]{tenoriotagle1981,lee1996}

\begin{equation}
    U_{\rm cloud}=-\frac{4}{3}\frac{v_{\rm impact}}{1+a}, \;\;\;
        U_{\rm disk}=\frac{4}{3}\frac{v_{\rm impact}}{1+a^{-1}}\,,
\label{eq:shock_velocities}
\end{equation}

\noindent where $v_{\rm impact}$ is the impact velocity of the cloud,  $a=\sqrt{n_{\rm cloud}/n_{\rm disk}}$, and $n_{\rm disk}$ is the disk density at the impact radius. \citet{muller2020} assumed the impact velocity equal to the Keplerian orbital velocity while the value adopted in this work is calculated as the vertical component predicted by the FRADO model, which results in only a fraction of the Keplerian speed. 

The characteristic timescale of the system is defined by the time required by the shock to cross the cloud. This assumption is true if the cloud is not previously destroyed by hydrodynamic instabilities or another process. Indeed, some authors have suggested that the BLR clouds have lifetimes of only a few months \citep{maiolino2010,ramosalmeida2017}. Nevertheless, these estimations consider that the clouds do not posses magnetic fields and their surrounding medium is stationary. Since clouds in the FRADO model are created from the accretion disk, it is more appropriate to think that they should be magnetized objects and therefore, survive longer \citep{shin2008}. We present a more detailed discussion about the stability of the clouds in Appendix \ref{sec:discussion}.

Shocks which are strong, adiabatic, and super-Alfv\'enic can accelerate particles efficiently up to relativistic energies by the Diffusive Shock Acceleration (DSA) mechanism \citep{bell1978,blandford1978}. In the case of parallel shocks and diffusion in the Bohm regime, the acceleration time for electrons and protons with energies $E_{e,p}$ in a magnetic field $B$ is \citep{drury1983, protheroe1999} 
\begin{equation}\label{eq:t_acc}
\frac{t_{\rm acc}}{\rm s} \sim 2.4\times10^6
\left(\frac{E_{e,p}}{\rm TeV}\right)
\left(\frac{B}{\rm G}\right)^{-1}
\left(\frac{v_{\rm sh}}{100\, \rm km\,s^{-1}}\right)^{-2}.
\end{equation}
Given the lack of direct information of the value of $B$ in these regions, we consider a sub-equipartition relation between the magnetic energy density and the ram pressure of the shocked cloud gas
\begin{equation}
    \frac{B^{2}}{8\pi}=\beta\,\frac{9}{8}m_p n_{\rm c}\,v_{\rm sh}^{2},
\end{equation}

\noindent where $\beta \le 1$ is the magnetization parameter. We obtain 
\begin{equation}\label{eq:B}
\frac{B}{\rm G} \sim 22
\left(\frac{\beta}{0.1}\right)^{\frac{1}{2}}
\left(\frac{n_{\rm c}}{10^{12} \rm cm}\right)^{\frac{1}{2}}
\left(\frac{v_{\rm sh}}{100\, \rm km\,s^{-1}}\right).
\end{equation}

\noindent If we adopt a conservative value of $\beta=0.1$, the typical magnetic field deduced for the inner clouds of the BLR is of hundreds of gauss. The estimated value becomes of the same order of magnitude assuming that the magnetic field is in sub-equipartition with the kinetic energy of the collision.

Particles escape from the acceleration region typically by diffusion, with timescale 
\begin{equation}\label{eq:t_diff}
\frac{t_{\rm diff}}{\rm s} \sim 5.1\times10^{3} 
\left(\frac{R_{\rm c}}{10^{12}\rm cm}\right)^2
\left(\frac{E_{e,p}}{\rm TeV}\right)^{-1}
\left(\frac{B}{\rm G}\right),
\end{equation}
in the Bohm diffusion regime. The adiabatic expansion of the perturbed cloud gas also decreases the energy of the particles, so that
\begin{equation}
 \frac{t_{\rm ad}}{\rm s}\sim 5\times10^5
 \left(\frac{R_{\rm c}}{10^{12}\,\rm cm}\right)\left(\frac{v_{\rm sh}}{100\,\rm km\,s^{-1}}\right)^{-1}.
 \label{eq:t_conv}
\end{equation}

\begin{figure} 
	\centering
	\includegraphics[width=\columnwidth]{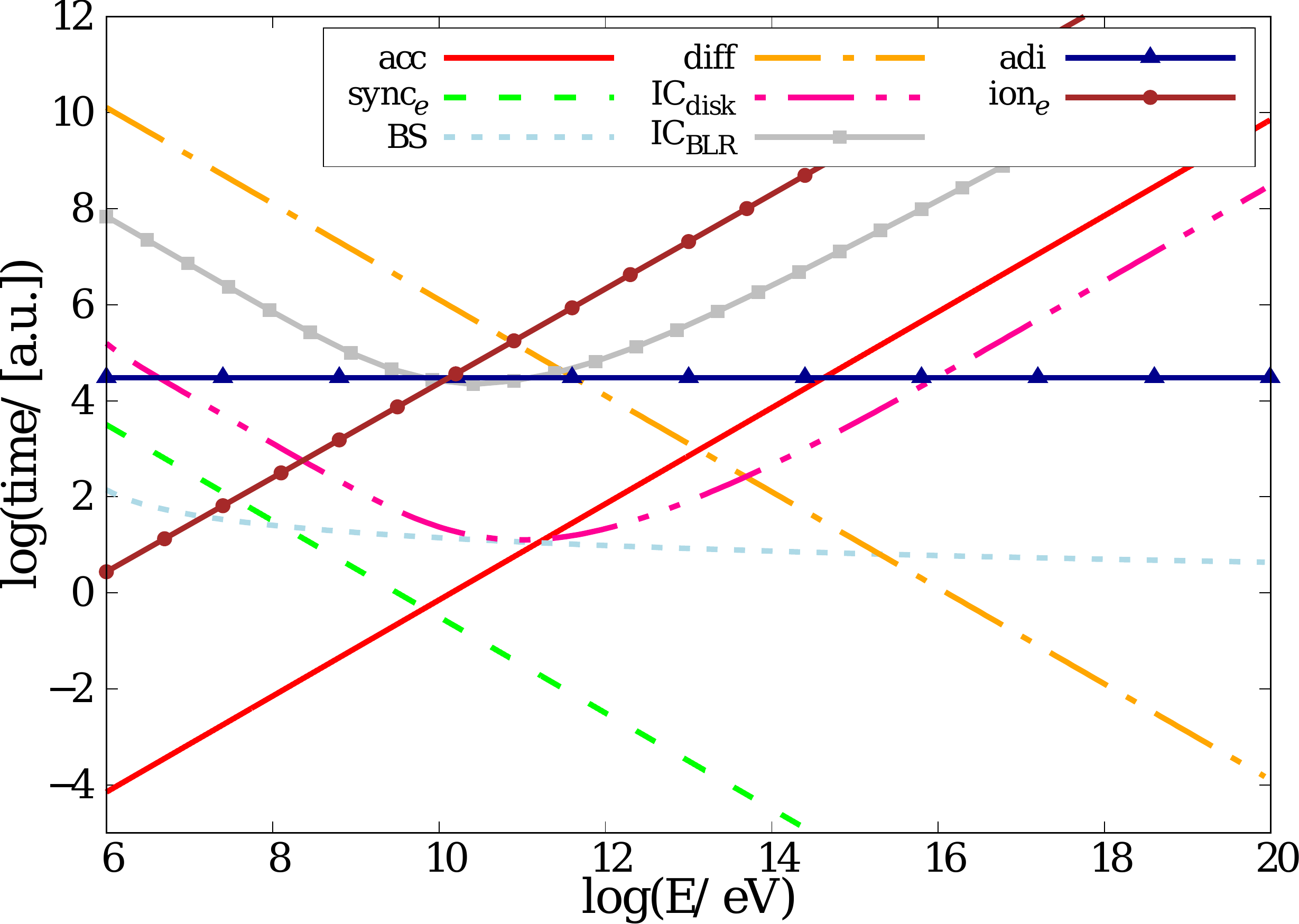}
	\caption{Acceleration, escape, and cooling timescales for electrons. This plot represents the typical relation between the different timescales for multiple parameter sets. For this reason, the y-axis is in arbitrary units.}
	\label{fig:cooling_e}
\end{figure}

\begin{figure} 
	\centering
	\includegraphics[width=\columnwidth]{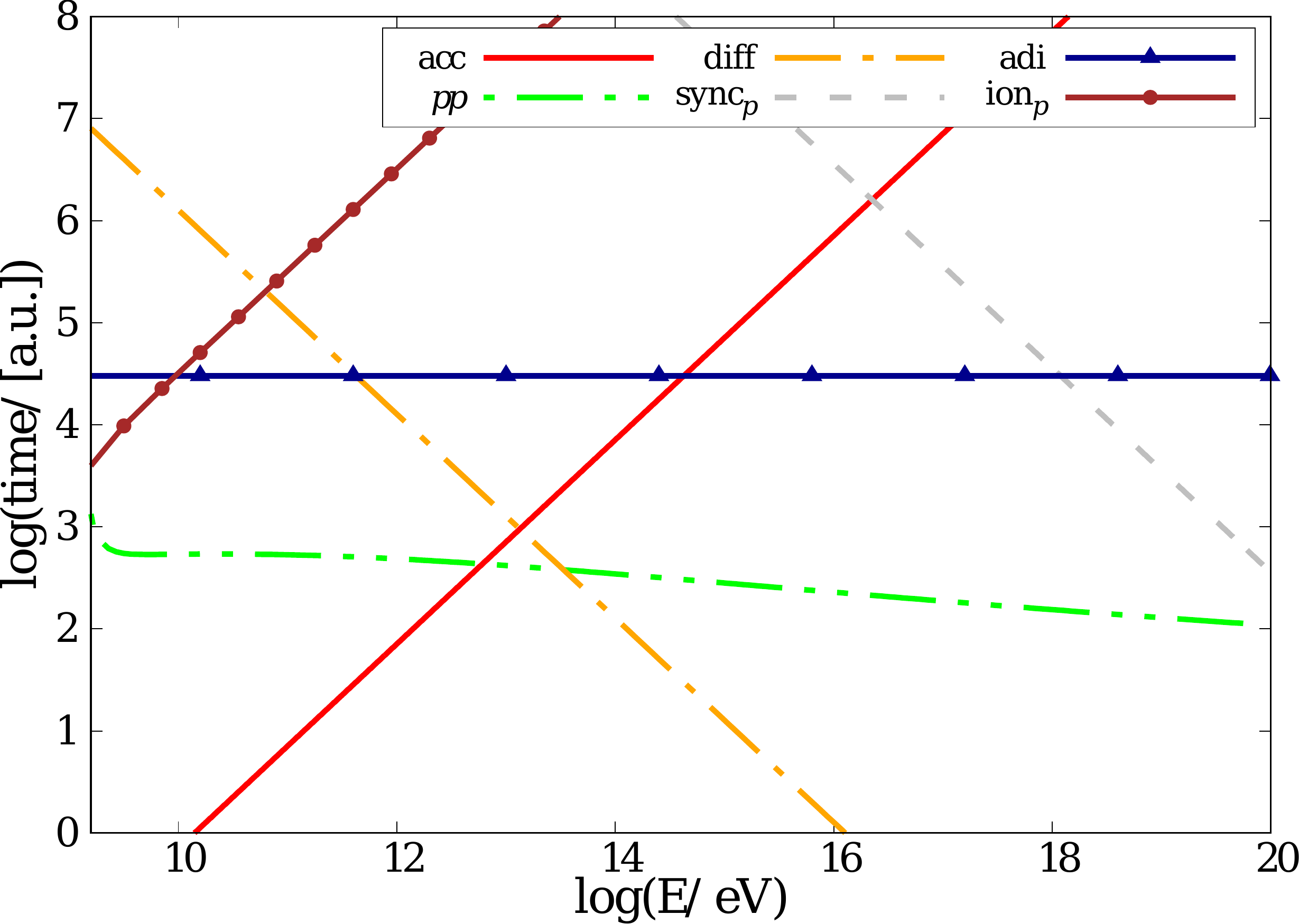}
	\caption{Idem to Fig.~\ref{fig:cooling_e}, but for protons.}
	\label{fig:cooling_p}
\end{figure}

The most relevant cooling mechanism for the electrons is the synchrotron one (see Fig.~\ref{fig:cooling_e}), operating at the timescale
\begin{equation}
    \frac{t_{\rm synchr}}{\rm s} \sim 
4\times 10^2
\left(\frac{E_e}{\rm TeV}\right)^{-1}
\left(\frac{B}{\rm G}\right)^{-2}.
\end{equation}
By equating $t_{\rm acc} = t_{\rm synchr}$ we find that the electrons maximum energy
constrained by synchrotron cooling is 
\begin{equation}\label{E_syn}
\frac{E_{\rm e, max}}{\rm GeV} 
\sim 17
\left(\frac{B}{\rm  G}\right)^{-\frac{1}{2}}
\left(\frac{v_{\rm sh}}{100\,\rm km\,s^{-1}}\right).
\end{equation}
The interaction of the electrons with the surrounding matter leads to relativistic Bremsstrahlung, whose characteristic timescale is given by
\begin{equation}\label{t_Brem}
\frac{t_{\rm Br}}{\rm s} \sim 2\times10^{3}
\left(\frac{n_{\rm c}}{10^{12}\,\rm cm^{-3}}\right)^{-1}. 
\end{equation}

Target photons for Inverse Compton scattering (IC) are provided by the BLR and the accretion disk, with energy densities $U_{\rm BLR}=0.1\,L_{\rm disk}/(\pi \,R_{\rm BLR}^{2}\,c)$ and $U_{\rm disk}=4\,\sigma_{\rm SB}\,T_{\rm disk}^{4}/c$, respectively (see Table \ref{tab:disk_parameters}). For computing the IC cooling rate and the emissivity, we follow the calculations presented by \citet{Aharonian1981}. The BLR and accretion disk photon fields are also taken into account for calculating the absorption of gamma-emission and the injection of secondary particles \citep{muller2020}. 

Protons are expected to interact with the local dense medium and lose energy due to proton-proton (pp) inelastic collisions on a timescale
\begin{equation}
    \frac{t_{\rm pp}}{\rm s} \sim 2\times10^{3} \left(\frac{n_{\rm c}}{10^{12}\,\rm cm^{-3}}\right)^{-1}.
\end{equation}
In Fig.~\ref{fig:cooling_p} we plot the relevant timescales for protons. We find that the maximum energy that protons can get is $E_{\rm p, max} \sim $~TeV and it is determined by the lifetime of the adiabatic shock.

We solve numerically the kinetic equation
\begin{equation}
  \frac{\partial N_{e,p}(E, t)}{\partial t}+\frac{\partial[b(E) N_{e,p}(E,t)]}{\partial E}+ \frac{N_{e,p}(E, t)}{t_{\textrm{esc}}(E)}=Q_{e,p}(E) \label{eq:particle_distribution}
\end{equation}
to find the energy distribution of electrons ($N_e$) and protons ($N_p$). In Eq.~(\ref{eq:particle_distribution}), $Q_{e,p} \propto E_{e,p}^{-2}$ is the injection rate of particles, $b(E) = E/t_{\rm cool}(E)$ accounts vice versa for the energy losses, and $t_{\rm esc}(E) = t_{\rm diff}(E)$. The luminosity due to injected relativistic particles is assumed to be $0.1\,L_{\rm s}$, where \mbox{$L_{\rm s}=\frac{1}{2}{M_{\rm c} v_{\rm impact}^{2}}/{t_{\rm coll}}$} is the power released during the collision \citep{delvalle2018}. Additionally, we consider this energy equally distributed between electrons and protons ($L_{\rm p}/L_{\rm e}=1$), as well as the case with hundred times more energy in protons than electrons ($L_{\rm p}/L_{\rm e}=100$). The lepto-hadronic spectral energy distribution (SED) are finally calculated using the expressions referred in \citet{muller2020}. 

\section{Results}\label{sec:results}

The efficiency of the production of non-thermal radiation from the shocks depends on the specific model parameters. The black hole mass, accretion rate, and metallicity directly affect the impact velocities of the clouds in the FRADO model, and subsequently, the shock properties. As a reference model, we use the solution with expected high non-thermal emission: $M_{\rm BH} = 10^8 M_{\odot} $, $\dot{m}=1$, and supersolar metallicity $Z=5 Z_{\odot}$, and we show the dependence of the results on these parameters.

\subsection{The Distribution of the Impact Properties and Shock Waves }

\begin{figure} 
	\centering
	\includegraphics[width=\columnwidth]{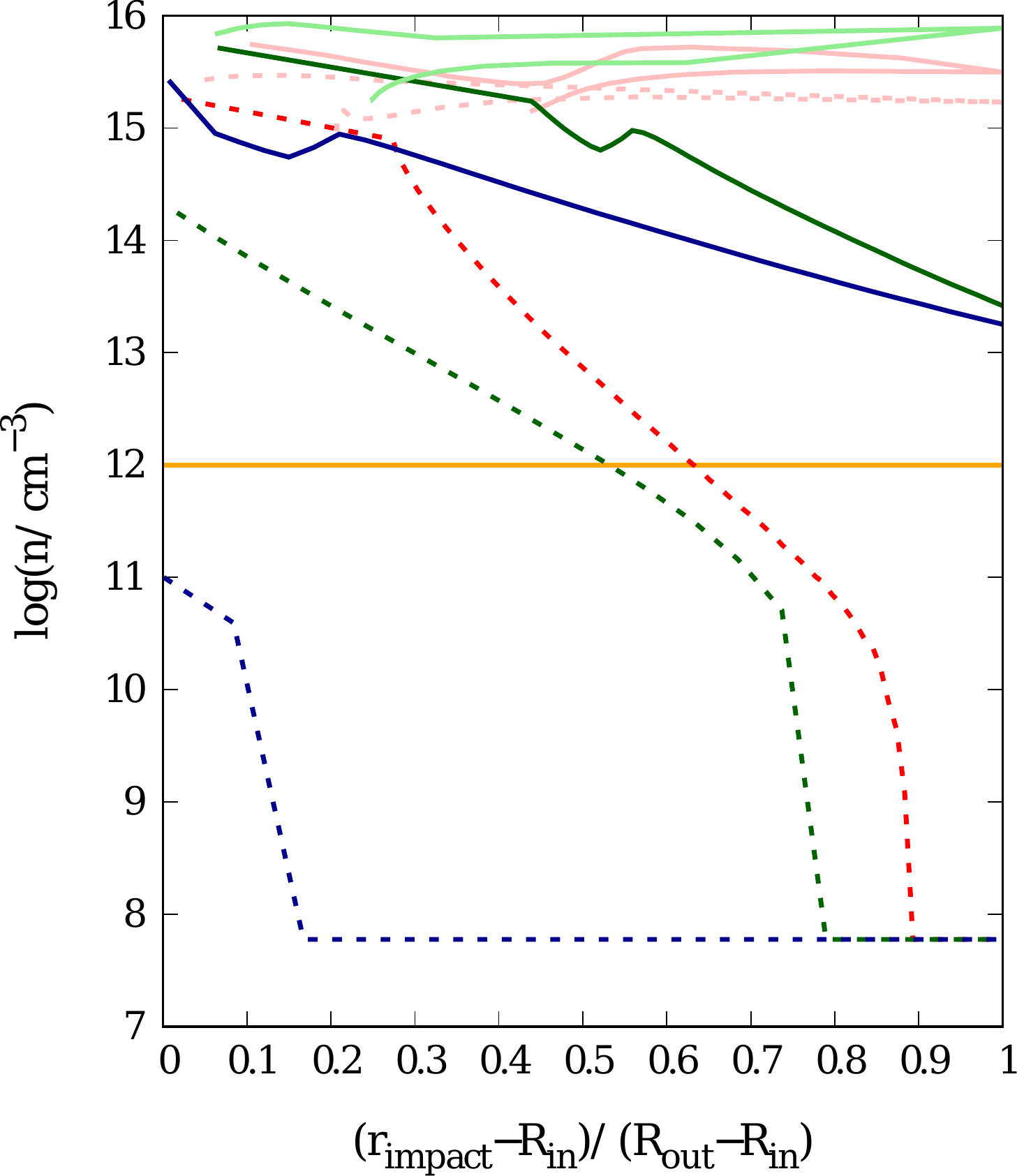}
	\caption{Density profiles of the unshocked disk and the unshocked cloud assuming Z=5 $Z_{\odot}$, $M_{\rm BH}=10^8$ (solid lines) and 10$^{6}$ M$_{\odot}$ (dashed lines), and $\dot{m}=0.01$ (blue), 0.1 (green), and 1 (red). The dark colors correspond to the local population of clouds, whereas the light colors are the non-local ones. The impact positions are normalized to the size of the BLR in each case, i.e., 0 is the position of BLR inner boundary R$_{\rm in}$, while 1 is the outer radius R$_{\rm out}$. The orange solid line indicates the density of the clouds, in this case fixed to $10^{12}$ cm$^{-3}$.}
	\label{fig:densities}
\end{figure}

Equations \ref{eq:shock_velocities} shows that the shock speed depends not only on the vertical component of the cloud velocity, but also of the cloud and disk densities. While the cloud density is fixed to $10^{11-12}$ cm$^{-3}$, the local density of the accretion disk is a function of $M_{\rm BH}$, $\dot{m}$, and $Z$, and the penetration depth of the cloud. In Fig.~\ref{fig:densities}, we display the density of the accretion disk at the location of the impact for the different parameter sets explored in this work. For a high black hole mass, the clouds penetrate much deeper into the disk, where the densities are higher, reaching some of them the equatorial plane. Such clouds, as mentioned in Section~\ref{sect:fall_back}, may cross the disk and even reappear on the other side. Nevertheless, since the crossing time of the clouds is much longer than the characteristic timescale of the system, we assume that in such cases the disk equatorial density determines the shock parameters. For small black hole mass the clouds are stopped very close to the disk surface. The dependence on the accretion rate is complex due to the fact that both the clouds velocities and the disk internal structure depend on this parameter.

\begin{figure} 
	\centering
	\includegraphics[width=\columnwidth]{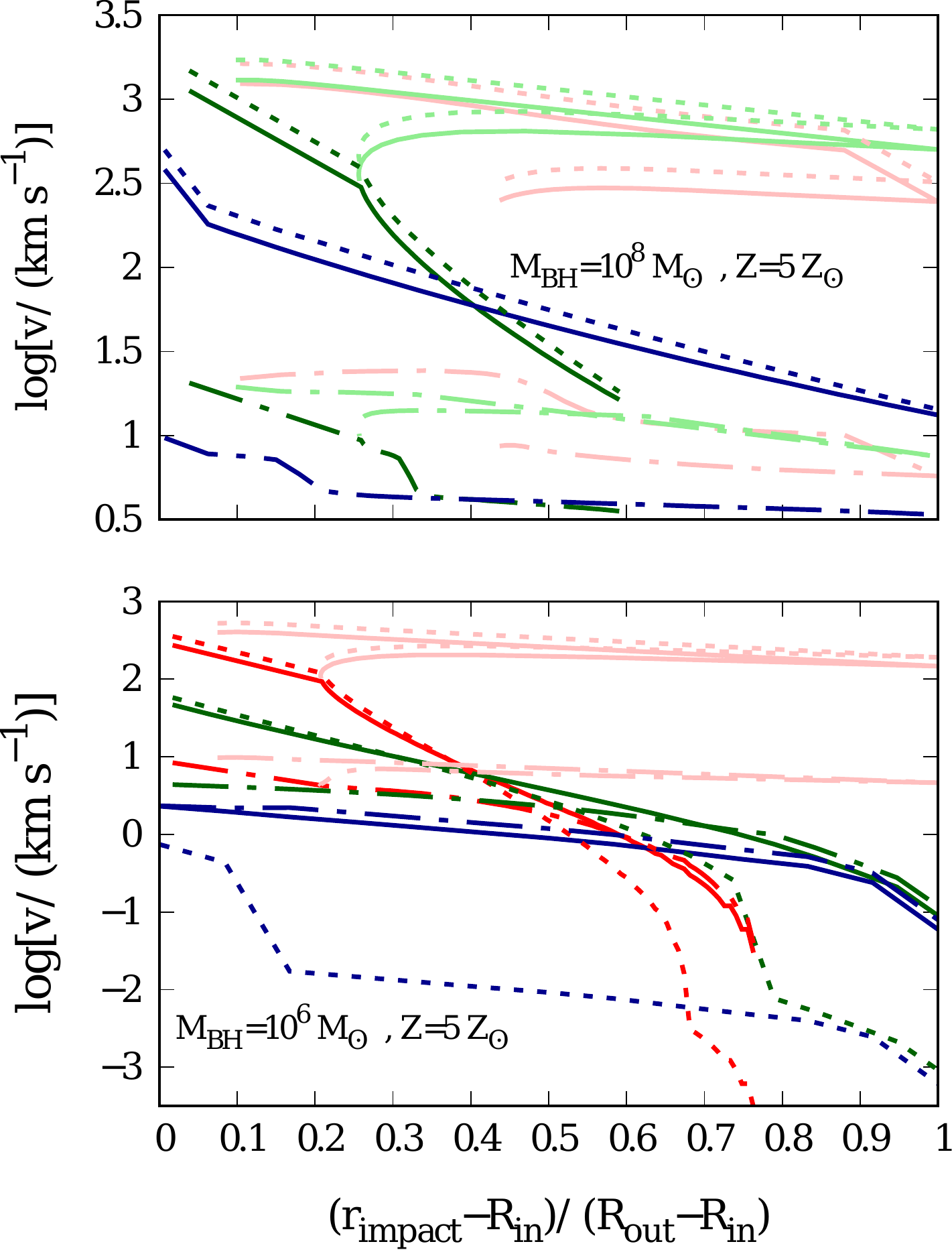}
	\caption{Impact (solid lines) and shock velocities assuming Z=5 $Z_{\odot}$, $M_{\rm BH}=10^8$ (top panel) and 10$^{6}$ M$_{\odot}$ (bottom panel), and $\dot{m}=0.01$ (blue), 0.1 (green), and 1 (red). The dotted lines show the velocity of the reverse shocks, whereas the dashed lines denote the velocity of the forward shocks. The dark colors are the local population of clouds, and the light colors the non-local. The impact positions are normalized to the size of the BLR.}
	\label{fig:shock_vel}
\end{figure}

The impact velocity of the clouds ($v_{\rm impact}$) and the associated shocks speeds ($U_{\rm disk}$, $U_{\rm cloud}$) are shown in Fig.~\ref{fig:shock_vel}. The impact velocity of the clouds increases with the accretion rate, the black hole mass, and the metallicity. In the very-low accretion regime, i.e., $\dot{m}=0.01$, the impact velocities of the clouds are very small in all the cases, creating very weak and slow shocks. On the other hand, in systems with a moderate- or high-accretion rate the clouds in the innermost part of the BLR might lead to the production of strong and fast reverse shocks. The latter can contribute to the non-thermal AGN component. In the next Section, we focus on systems with strong and fast shocks and model their expected non-thermal features.

\subsection{Spectral Energy Distributions}

\begin{figure} 
	\centering
	\includegraphics[width=\columnwidth]{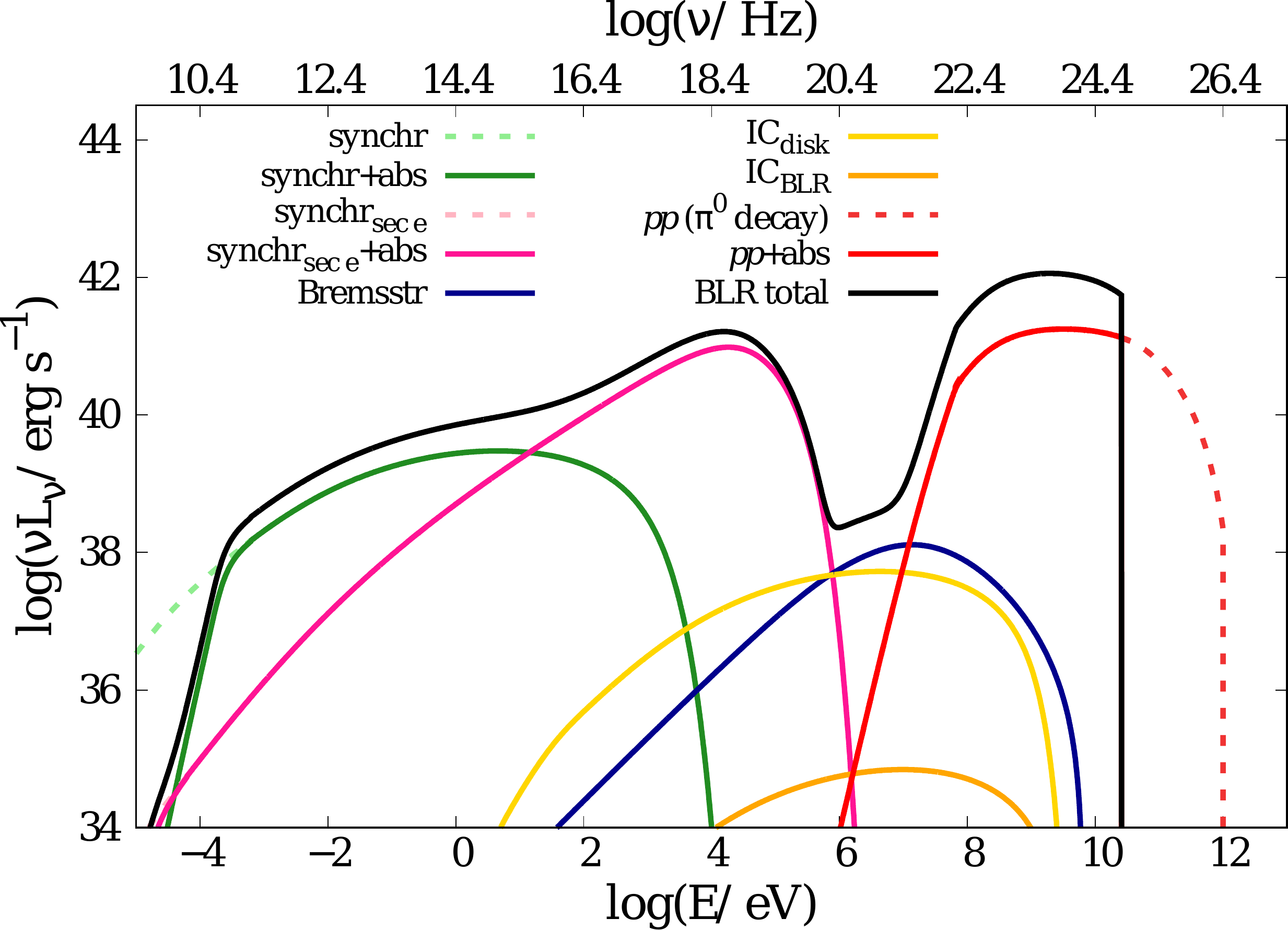}
	\caption{Spectral energy distribution for electrons and protons in a special case of $Z=5 Z_{\odot}$, $M_{\rm BH}=10^{8}$ M$_{\odot}$, $\dot{m}=1$, $n_{\rm c}=10^{12}$ cm$^{-3}$, $R_{\rm c}=10^{12}$ cm, and a ratio of energy injected in protons to electrons $L_{p}/L_{e}=100$. The black line shows the total BLR emission, whereas the other lines represent the contribution from the most inner ring of clouds.}
	\label{fig:sed}
\end{figure}

In Fig.~\ref{fig:sed} we plot the SED for the particular case with $Z=5 Z_{\odot}$, $M_{\rm BH}=10^{8}$ M$_{\odot}$, $\dot{m}=1$, $n_{\rm c}=10^{12}$ cm$^{-3}$, $R_{\rm c}=10^{12}$ cm, and a ratio of energy injected in protons to electrons of $L_{\rm p}/L_{\rm e}=100$. We find that the non-thermal emission at low frequencies is dominated by synchrotron radiation of the primary electrons, whereas the X-ray is dominated by the synchrotron radiation of the secondary pairs injected by the absorption of gamma-ray photons in the BLR and accretion disk photon fields \citep[see][for more details about the secondary electrons]{muller2020}. The emission at higher energies is dominated by the gamma-rays coming from the decay of neutral pions produced in proton-proton inelastic collisions. The main difference with the results obtained in the previous work is the drop of the luminosity observed around MeV, which is a consequence of the $3-4$ orders of magnitude smaller maximum energy that protons can achieve with the orbits of the clouds described by the FRADO model, because of the lower effective impact velocities. 

\begin{figure} 
	\centering
	\includegraphics[width=\columnwidth]{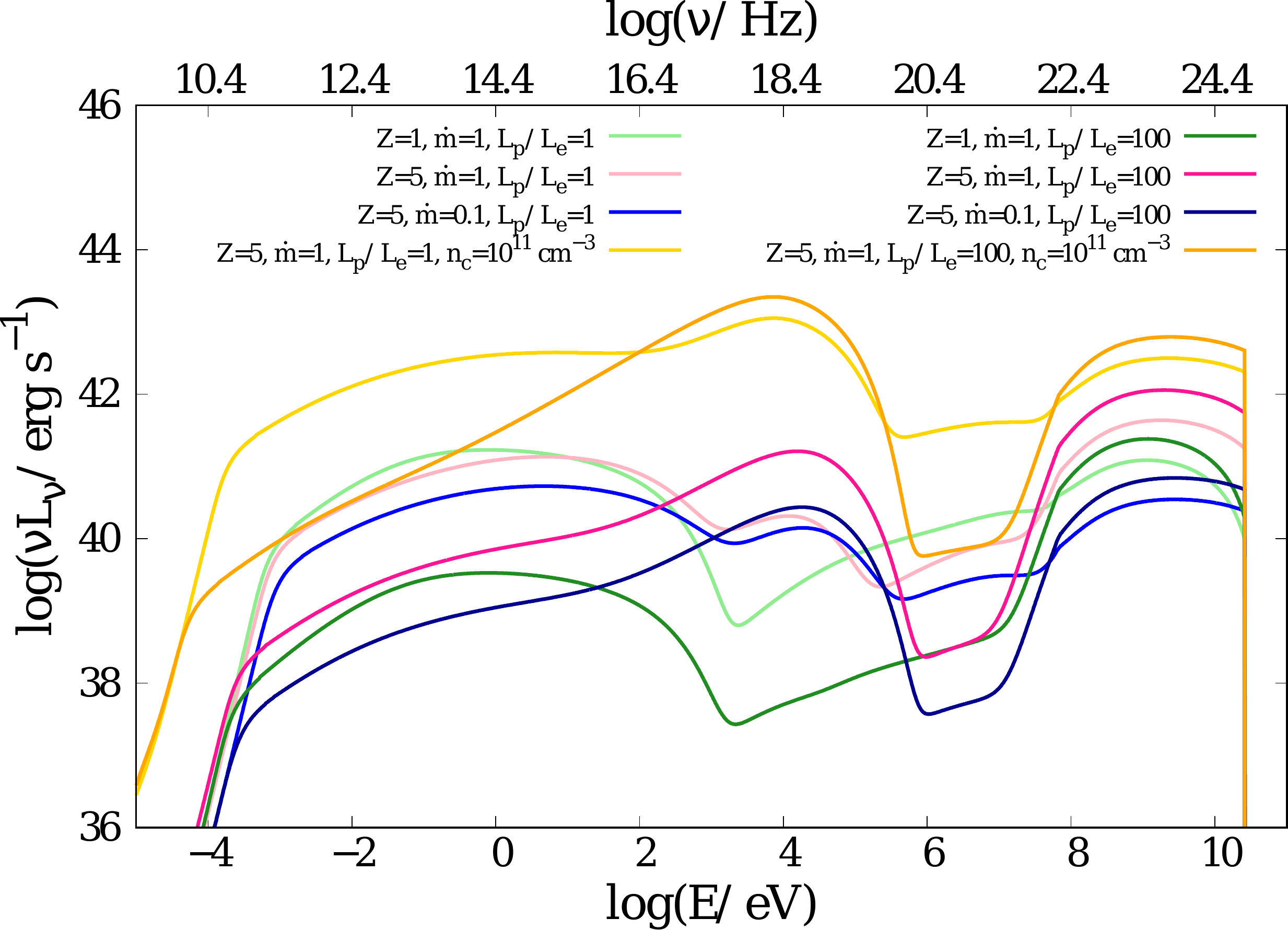}
	\caption{Spectral energy distribution for different parameter sets. In all the cases the mass of the supermassive BH is $10^{8}$ M$_{\odot}$.}
	\label{fig:sed_totals}
\end{figure}

Figure \ref{fig:sed_totals} shows the total emission integrated along the BLR for the different parameter sets. It is possible to see that the accretion rate does not modify the shape of the SED, but the luminosity increases with its value. On the other hand, as lower metallicity values implies smaller impact velocities and smaller maximum energies for the particles, this affects the injection of secondaries and the SEDs obtained for different metallicities differ in the slope in the X-ray/soft-gamma-ray ranges. 

We also observe that one order of magnitude difference in the density of the clouds produces a large difference in the luminosities but not in the shape of the SED. This is because the velocity of the shock wave in the less dense medium increases and the lifetime of the adiabatic shock is prolonged as well. Therefore, the fraction of energy released in the impact that can be injected in relativistic particles rises. Thus, lower density BLR clouds ($10^{11}$ cm$^{-3}$) combined with high black hole mass, high accretion rate, and high metallicity bring the non-thermal emissivity level to values $2\times10^{43}$ erg s$^{-1}$, in the X-ray band, peaking around $\sim 10$ keV. The spectral shape is additionally sensitive to the adopted energy injection as shared between electrons and protons. Nevertheless, this ratio could be constrained with radio observations. Furthermore, future comparison with broad band X-ray/gamma-ray data will allow to determine whether this mechanism can indeed contribute significantly to the hard X-ray emission of non-jetted sources. It can also help to constrain additionally the parameters in BLR models, specifically in the FRADO model.

\begin{figure} 
	\centering
	\includegraphics[width=\columnwidth]{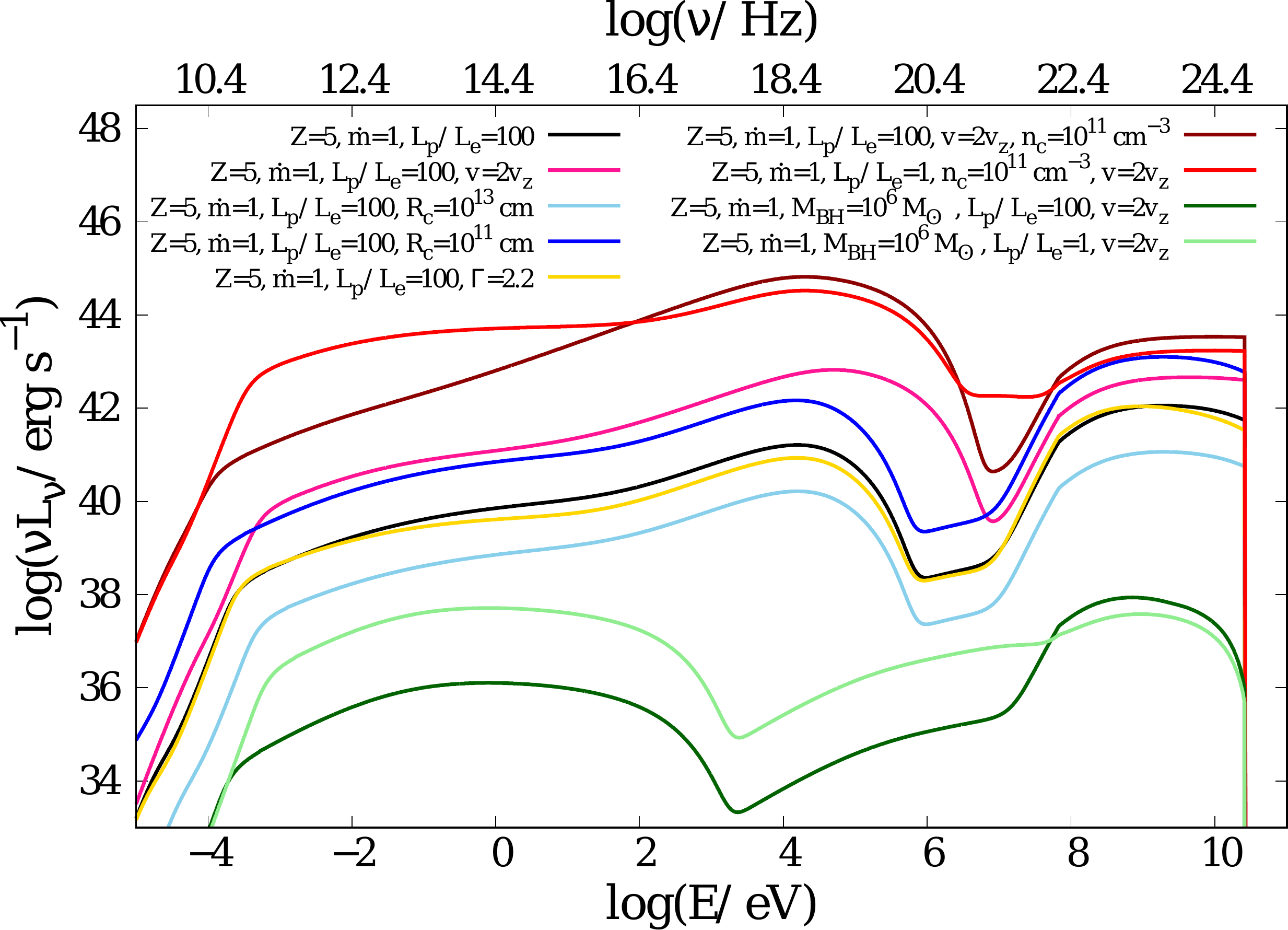}
	\caption{Spectral energy distribution for different parameter sets. The model with $Z=5\,Z_{\odot}$, $M_{\rm BH}=10^{8}$ M$_{\odot}$, $\dot{m}=1$, $R_{\rm c}=10^{12}$ cm, $n_{\rm c}=10^{12}$ cm$^{-3}$, and $L_{p}/L_{e}=100$ is plotted as the reference model.}
	\label{fig:sed_comparison}
\end{figure}

As we mentioned before, considering the impact velocity equal to the vertical velocity component produces a drop in the spectra at MeV. In Fig.~\ref{fig:sed_comparison}, we show how the results change if we duplicate the effective impact velocity, change the cloud radius, or the particle injection index. We found that the total luminosity evolves inversely proportional to the cloud size. On the other hand, injecting relativistic particles following a power law with index $2.2$ modifies the shape of the SED softening the gamma-ray spectrum and, consequently, reducing the X-ray emission. Finally, if the effective velocity for the cloud impact is assumed to be twice its vertical component, the maximum energy of the particles increases ($\sim 10$ TeV), and the gamma spectrum becomes harder. Therefore, the X-ray emission rises and extends up to higher energies, filling the gap in the MeV emission observed for lower effective impact velocities. In addition, systems with less massive black holes ($10^{6}$ M$_\odot$) starts to radiate significantly. If the effective velocity matches the Keplerian velocity, the results obtained by \citet{muller2020} are reproduced. Nevertheless, to know how much kinetic energy can exactly be transferred to the production of shock waves is a challenging task and it will require us to perform careful magnetohydrodynamic simulations in the future. 

\subsection{Comparison of the Predicted X-ray Non-Thermal Flux with the Data for AGN}

A simple analysis of the predicted role of the non-thermal emission in non-jetted sources can be done on the basis of the observed UV/X-ray statistical trend \citep{lusso2016}. Fits to the UV flux measured at $2500$ \AA ($L_{UV}$) and 2 keV ($L_X$) gives a non-linear relation between the two
\begin{equation}
\log L_X = 27.46 + 0.53 \log L_{UV},
\label{eq:UV-X}
\end{equation}
\noindent where the exact values were taken from \citet{salvestrini2019}. Our model shown in Fig.~\ref{fig:sed_totals} has the bolometric luminosity of $\sim 10^{46}$ erg s$^{-1}$. Using the bolometric correction of 27 to obtain the 2 keV flux we expect the X-ray flux there at a level of $4 \times 10^{44}$ erg s$^{-1}$. The model with a high cloud density predicts the flux there to be $\sim 10^{41}$ erg s$^{-1}$, suggesting a few per cent contribution to the total X-ray flux. However, the low cloud density predicts a 2 keV flux of $\sim 10^{43}$ erg s$^{-1}$, which combined with a higher effective impact velocity could lead to a 2 keV luminosity of $\sim 10^{44}$ erg s$^{-1}$, as shown in Fig.~\ref{fig:sed_comparison}.

\citet{Laor2008} have found that, in radio-quiet AGNs, the luminosity at 5 GHz and the integrated X-ray luminosity ranging from 0.2 to 20 keV follows the relation $L_{5 {\rm GHz}}/L_{0.2-20 {\rm keV}} \sim 10^{-5}$. The exact SED shape of our models in the low-energy radio domain depends on the accretion rate as well as the ratio $L_{p}/L_{e}$. In the cases with the higher X-ray emission, i.e., $M=10^{8}$ M$_{\odot}$, $Z=5\,Z_{\odot}$, $n_{\rm c}=10^{11}$ cm$^{-3}$, and $L_{p}/L_{e}=100$ (see Figs.~\ref{fig:sed_totals} and \ref{fig:sed_comparison}), we find that $L_{5\,{\rm GHz}}/L_{0.2-20\,{\rm keV}} \sim 10^{-7}$ as a consequence of the synchrotron self-absorbed emission at frequencies below $\sim 10$ GHz. This result indicates that if the interaction of BLR clouds with the accretion disk contributes significantly to the X-ray band, an additional source emitting at 5 GHz is needed in order to match the trend observed by \citet{Laor2008}. \citet{Behar2015,Behar2018} evaluate an analogous relation using a higher radio frequency. Their study indicates that $L_{95\,{\rm GHz}}/ L_{2-10\,{\rm keV}}=10^{-4}$. In our case, we find that $L_{95\,{\rm GHz}}/ L_{2-10\,{\rm keV}} \approx 1.5\times10^{-4}$. This shows that the non-thermal emission has a potential to contribute to the observed X-ray flux in non-jetted sources but the model predictions are extremely sensitive to the adopted cloud density. Therefore, future comparison with the observational data has to concentrate more on the specific shape of the broad-band X-ray spectrum. 

We stress that while we consider a contribution of the non-thermal emission to X-ray spectra we do not claim that non-thermal emission produced via cloud-disk interactions is the sole mechanism of X-ray radiation in radio-quiet sources since that would contradict the presence of the relativistic Compton reflection, relativistically broadened iron K$\alpha$ line, and all energy-resolved time delay studies \citep[e.g.,][]{pounds1990,reynolds1999,dovciak2004,2021bhns.confE...1K,2022AdSpR..69..448K}.

In particular, this model cannot reproduce the characteristic short-scale X-ray variability observed in AGNs. The stochastic variability level in our model is low, since the total number of clouds in BLR is high, and even the number of impacts per second is of order of 6000 for a model with 5 $Z_{\odot}$, $M_{\rm BH} = 10^8 M_{\odot}$, and $\dot{m}=1$, which implies a variability level of 1 \%. However, the variations of the accretion rate in the innermost part of the flow can be very fast, as inferred from studies of some Changing Look AGN \citep[e.g.,][]{ngc1566-2022}. Such a sudden change of the bolometric luminosity affects the motion of the clouds high above the disk and modifies the impact radii and velocities. In addition, observations of stellar winds in supergiant stars clearly show that wind launching is episodic \citep[e.g.,][]{humphreys2022}. Such a type of outflow behavior can additionally enhance the level of variability.

\subsection{Comparison of the Predicted Gamma-Ray Flux with the Data for AGN}

Despite the fact that AGN make up at least 80\% of the gamma-ray resolved sources, the number of identified GeV radio-quiet/radio-weak objects is just a few tens \citep{Fermi4th}. For this reason, a relation similar to Eq.~(\ref{eq:UV-X}) has not been established yet, although some authors have already suggested that the gamma-emission seems to correlate with the X-ray radiation, and therefore, with the properties of the AGN \citep{Wojaczynski2015,Wojaczynski2017}. Within the group of potential weak-jetted gamma-ray AGN, there is one particular kind of object which has been growing in popularity and number: the so-called narrow-line Seyfert 1 galaxies \citep[NLSy1s;][]{Paliya2019}. The peculiar spectra of these galaxies show the broad-emission lines observed in Seyfert 1 galaxies, but narrower, and also a steep gamma-ray fall similar to flat-spectrum radio quasars. Therefore, it has been proposed that they represent an intermediate state between these two AGN classes. Since Seyfert galaxies are high-accreting objects, they become natural candidates for the model presented in this work. As we showed, the efficiency of the production of the non-thermal emission by the collisions of BLR clouds increases with the accretion rate. In \citet{muller2020}, the authors have already found that if the energy transmitted to the production of the shock waves corresponds to the Keplerian velocity of the clouds, the model explains successfully the gamma-emission from the nearby Seyfert galaxy NGC 1068. Nevertheless, the detection of more distant galaxies is very challenging because of the sensitivity of the current gamma-ray instruments.

The observational information of NLSy1s seems to indicate that the gamma-emitter objects are radio-loud sources containing a pc-scale jet with Doppler factors of $5-10$, and therefore, the GeV radiation is typically associated with the existence of a jet. Rather indirect argument for a jet-powered non-thermal radio emission comes from the first detection of the fast (day) variability in NLSy1 Mrk 110 \citep{panessa2021}, however, the compactness of the region is inferred from the timescale while it might reflect a turbulence at somewhat larger radii. In any case, the radio and gamma-ray emission do not need to have the same origin. Additionally, not all the radio-loud NLSy1s have been detected at gamma-rays energies.

\citet{Paliya+2019} compiled the data of all the discovered gamma-NLSy1s, which are only 16 (only 9 of them are bona-fide sources according to \citet{DAmmando2019}). The authors found that the fit of the SEDs requires to assume that the contribution of the components is not the same in every source, e.g., the X-ray radiation is sometimes explained by a corona and synchrotron self-Compton radiation and other times by the Compton up-scattering of external photons. On the other hand, the gamma-ray emission is always explained by inverse Compton, but sometimes with photons from the BLR and other times from the dusty torus. Moreover, GeV blazar-like flares of four gamma-NLSy1s have been reported over the last years \citep{Gokus2021}. Some of the flares are simultaneously observed in soft X-ray and gamma-rays, but not all of them, and non-hard X-ray variations have been detected, suggesting that there should exist several emission regions \citep{Paliya2019}. In particular, SBS 0846+513 and PKS 1502+036 have been observer in gamma-rays during the low-state without showing any coincidence with superluminal motion in radio. In the case of SBS 0846+513, superluminal motion has been also measured without an associated gamma-ray flare. Furthermore, the low- and high-activity SEDs can be modelled as synchrotron and external Compton radiation, but the transition between both states requires not only a change in the electron distribution of the emitting region, but also a change in the magnetic field value \citep{DAmmando2019}. This can also be interpreted as a different radiative region dominating the gamma-emission during the different states, as observed in the X-ray band of some NLSy1s where the corona prevails during the low gamma-ray activity and the jet during the gamma-ray flares \citep{Paliya2019}.

The model presented in this work does not aim to explain flaring episodes of a few hours, which are likely related to the presence of a relativistic outflow, but it could significantly contribute to the continuous and/or long-variable emission from NLSy1s. During the flares, electrons accelerated in the jet might dominate the gamma-ray produced by external Compton scattering of photons, whereas in the low-states the radiation could mainly be provided by other components, e.g., the collisions in the BLR. As indicated in Fig.~\ref{fig:sed_comparison}, the luminosity expected from our model rises with the mass of the central BH and the accretion rate. Therefore, gamma-rays of distant NLSy1s with typical black holes of $10^{8}$ M$_{\odot}$ are detected, in contrast to nearby Seyfert galaxies with typical masses of $10^{6}$ M$_{\odot}$. In addition, our model can reproduce the photon index of $\sim 1.6$ that is commonly observed at hard X-ray energies \citep{Paliya2019} and offers a hadronic contribution to the gamma-emission of NLSy1s, which leaves open the possibility of neutrino production.

In radio-quiet AGN as well as quiescent galactic nuclei such as the Galactic Center, another potential source of the long-term gamma-ray emission is a population of old millisecond pulsars. It is expected that there is a substantial population of neutron stars of various ages in the dense nuclear star clusters (NSCs), such as inside the NSC in the Galactic Center that has been studied in detail \citep{2015AcPol..55..203Z,2017FoPh...47..553E,2017CoSka..47..124K, 2017A&A...602A.121Z}. In particular, the sub-population of old, accretion-powered millisecond pulsars could be a source of the extended and diffuse gamma-ray emission as analyzed by \citet{2005MNRAS.358..263W} and recently by \citet{2021arXiv210600222G}. In this scenario, the gamma-rays are produced by the prompt emission of electron-positron pairs in the millisecond pulsar magnetospheres and also by the delayed emission of the escaping pairs in the pulsar wind shocks via the inverse Compton scattering on the interstellar radiation field. In both cases, the gamma-ray emission originates in the rotational kinetic energy of the pulsars at the age of $\sim 13.8$ Gyr via the magnetic dipole braking. The model has been successful in explaining the Galactic center gamma-ray excess \citep{2021arXiv210600222G}. The peak luminosity that possibly originates from a millisecond pulsar population is $L_{\rm MSP} \sim 4\times  10^{36}\,{\rm erg\,s^{-1}}$ (see Fig.~3 in \citeauthor{2021arXiv210600222G}, \citeyear{2021arXiv210600222G}), which is two orders of magnitude smaller in comparison to the non-thermal GeV luminosity from cloud-disk collisions (for $10^6\,M_{\odot}$ BH, see Fig.~\ref{fig:sed_comparison} in this paper, where the GeV luminosity is $\sim 10^{38}\,{\rm erg\,s^{-1}}$). Hence the GeV emission of millisecond pulsars, in case there are present in radio-quiet AGN, is not expected to preclude the detection of the BLR non-thermal GeV emission.   

\subsection{Model Self-Consistent Accretion Rate}

\begin{figure} 
	\centering
	\includegraphics[trim={0.25cm 0.25cm 0.25cm 0.25cm},clip, width=\columnwidth]{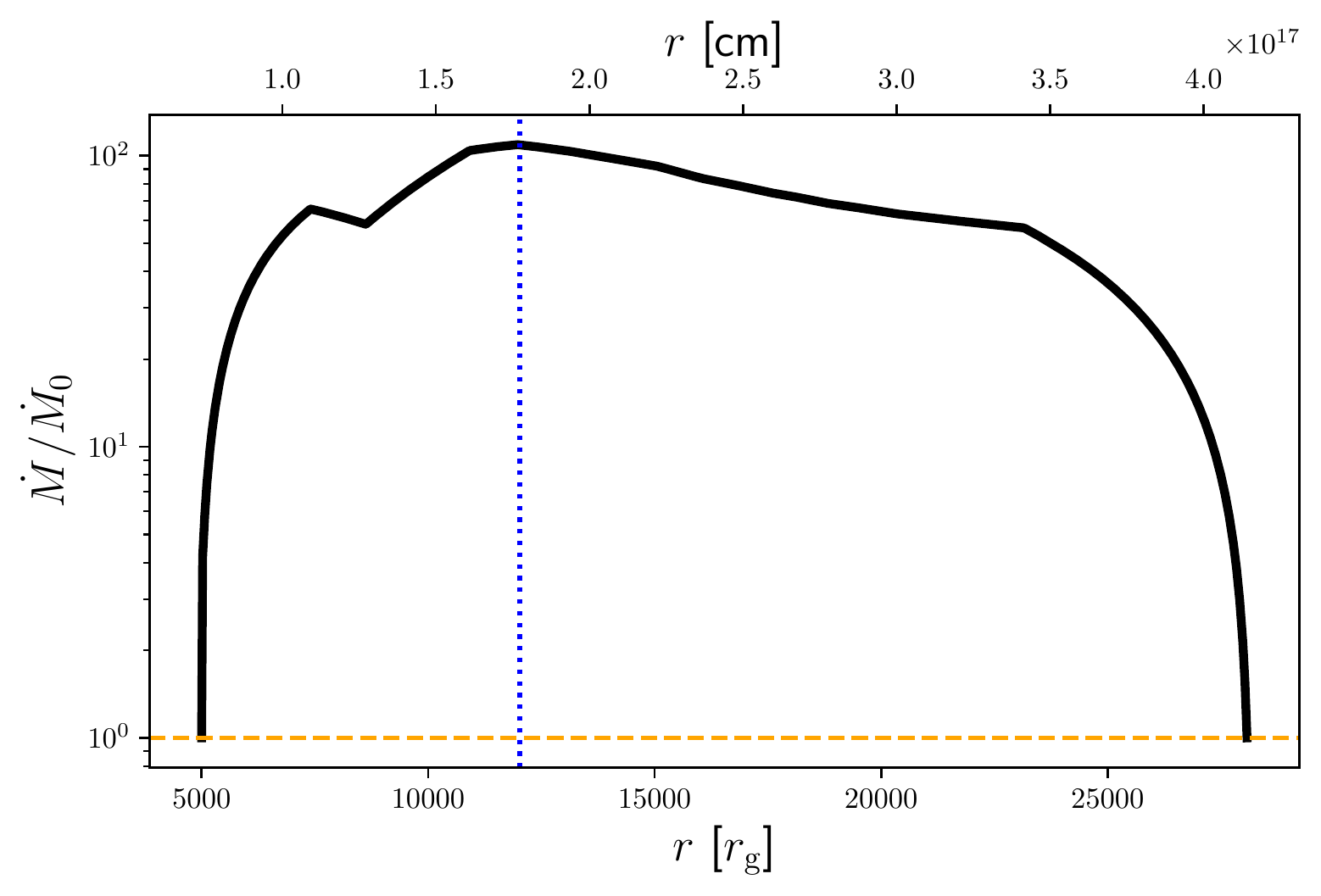}
	\caption{The ratio of the local accretion rate to the initially adopted disk accretion rate for the most extreme case with initial parameters of $M_{\rm BH}=10^{8} M_{\odot}$, $Z=5\,Z_{\odot}$, $\dot{m}=1$, and $n_{\rm c}=10^{11}$ cm$^{-3}$. The local accretion rate was calculated by solving the continuity equation including local net effect of the infall and outflow of the clouds from the disk surface. Inflow dominates at large radii so the net accretion rate rises initially towards smaller radii, reaches the maximum at $12\,023\,r_{\rm g}$ (marked by blue dotted line), and later returns to the initial value due to outflow. This effect is not included in the disk computations.}
	\label{fig:accretion_rate}
\end{figure}

In this work, we consider that the accretion rate is constant along the accretion disk in order to save computing time. Nevertheless, the stream of matter created by the infalling and outflowing clouds can lead to local variations of the accretion rate along the accretion disk radius. In Fig. \ref{fig:accretion_rate}, we display an estimation of the net accretion rate taking into account the outflow/inflow of clouds for the model parameters corresponding to the most efficient production of non-thermal emission. This result shows that the stream of matter produces a local rise of the accretion rate along the BLR length. For a system with $M_{\rm BH}=10^{8}$ M$_{\odot}$, $Z=5\,Z_{\odot}$, $\dot{m}=1$, and $n_{\rm c}=10^{11}$ cm$^{-3}$, we find a maximum at $r \approx 1.2\times10^{4}\,r_{\rm g}$, where the accretion rate reaches a value two orders of magnitude above the model value. To account for this effect, it will be required to perform iterative calculations of the disk structure and the BLR dynamics which are out of the scope of this paper, but which should be implemented in the future. The outcome of such computations is not simple to evaluate. Much larger accretion rate in the efficient impact zone would imply still higher outflow and inflow rates, as can be inferred from Eq.~\eqref{eq:impact}. However, the effective temperature in the involved region rises in this case above the sublimation temperature of the dust which actually would turn off the outflow completely in the zone, moving the outflow zone outwards. A stationary solution could form, with the outflow zone slightly more distant than in the model without self-consistency requirement satisfied. In that case, the outflow would be less vigorous since the central flux available for the material high above the disk will not be enhanced (inner accretion rate is fixed) and may be slightly suppressed due to the somewhat larger distance of the launching zone. The locally enhanced continuum emission will contribute to the near-IR and may actually mimic the warm dust component in the spectrum. On the other hand, a non-stationary solution could lead to an oscillating position of the outflow zone, which may produce time-dependent behaviors in the spectrum. 

\section{Summary and conclusions} \label{sec:conclusions}

In this paper, we expand the study by \citet{muller2020} of the non-thermal emission arising from BLR cloud impacts onto the accretion disk adopting the FRADO model for a detailed description of the clouds motion and computing the accretion disk internal density distribution. In this model clouds are launched initially by the local radiation density field, but they are accelerated after by the non-local radiation coming from the unshielded part of the accretion disk. Cloud velocities are thus much higher than expected just from the locally available energy dissipated in accretion disk and available as radiation flux but their vertical components are nevertheless lower than the Keplerian velocities adopted by \citet{muller2020}.

We show that reduced impact velocities in comparison with the original work lead to an efficient production of non-thermal emission, but only for a fraction of the parameter space. The assumption of solar metallicity of the material does not lead to impact velocities large enough to produce non-thermal emission efficiently. Nevertheless, considering the metallicity five times higher enhances the predicted flux by two orders of magnitude. This fact agrees with the results of \citet{naddaf2021} and \citet{naddaf2021b} where more powerful outflows are produced at higher metallicities. This implies that in case there is non-thermal emission associated to the dynamics of the BLR, the metallicity and the intensity of the radiation will be linked. The second key parameter of the model is the cloud density. Denser cloud models ($n_c \sim 10^{12}$ cm$^{-3}$) under-predict the expected X-ray luminosity at 2 keV in non-jetted AGNs by several orders of magnitude, while lower density clouds ($n_c \sim 10^{11}$ cm$^{-3}$) could contribute to the X-ray emission. Interestingly, not only the luminosity but also the shape of the non-thermal emission depends on the Eddington rate of the source, which manifests that the intrinsic properties of the AGN are imprinted in the radiation. If the energy fraction released in the impact is at least twice the fraction of the kinetic energy calculated using only the vertical component of the velocity computed with the FRADO model, the SEDs present some similarities to the ones observed in gamma-NLSy1s. Nevertheless, determining precisely the effective impact velocities will require detailed magnetohydrodynamic studies. In all the cases, the spectra show pronounced differences in the MeV range. 

Our model predicting the cloud-disk interaction is based on the assumption that clouds are launched and accelerated solely by the radiation pressure acting on dust. However, higher above the disk, the line-driving mechanism certainly contributes to the acceleration as well, and the final cloud velocities might be larger. At present, we cannot calculate quantitatively the combined effect of the radiation and line-driving mechanisms since this is a very complicated computational problem. However, in the future, the dust-based code should be combined with line-driving codes, such as QWIND \citep{QWIND2010,QWIND32021}.

The non-hydrodynamical models use the concept of clouds instead of a continuous wind. In reality, the outflow is probably launched as a roughly uniform wind, and later instabilities lead both to formation of the clouds embedded in the intercloud (almost co-moving) medium, and multiple processes work towards cloud condensation and cloud destruction. These processes are observed in stellar winds. We discuss these unresolved issues extensively in Appendix \ref{sec:discussion}.

The scenario of cloud-disk interaction presented in this paper could be enriched by considering additional processes. For instance, the bow shocks created by the clouds moving with Keplerian velocities through the surrounded medium could also accelerate particles and be an extra source of non-thermal emission. On the other hand, complete dynamical models as FRADO can provide the statistics of clouds which escape and do not collide with the disk. This can be employed to estimate the probability of clouds that enter jets in jetted systems \citep[see e.g.,][]{dar1997,araudo2010,delpalacio2019}. In addition, considering BLR models that describe the photon field at each point inside these regions, can improve the non-thermal emission and absorption maps at the highest energies. Therefore, the observations performed by the next generation of MeV and gamma observatories, such as \textit{AMEGO} or \textit{AMEGO-X} and the Cherenkov Telescope Array, will contribute substantially to these studies \citep{AMEGO,CTA}.

\section*{acknowledgments}
We are grateful to the anonymous reviewer for the constructive comments that helped us to improve the manuscript. A.L.M., A.T.A. and V.K. thank the Czech Science Foundation under the grant GA\v{C}R 20-19854S titled ``Particle Acceleration Studies in Astrophysical Jets''. The project was partially supported by the Polish Funding Agency National Science Centre, project 2017/26/A/ST9/00756 (MAESTRO 9), and MNiSW grant DIR/WK/2018/12. The authors also acknowledge the Czech-Polish mobility programs (M\v{S}MT 8J20PL037 and
NAWA PPN/BCZ/2019/1/00069). MZ acknowledges the financial support by the GA\v{C}R EXPRO grant No. 21-13491X ``Exploring the Hot Universe and Understanding Cosmic Feedback''.  

\appendix

\section{Cloud formation and destruction, shock formation mechanisms}\label{sec:discussion}

The presented scenario of non-thermal emission is based on the assumption of the cloud formation soon after the wind is launched, and cloud survival until the moment of the impact onto the disk surface. The formation and existence of the two-phase medium at a BLR distance is a difficult topic although the possibility of the thermal instabilities in the irradiated environment in AGN has been already considered several decades ago \citep{1981ApJ...249..422K,begelman1990,2000MNRAS.316..473R}. However, the stability of cloud structures close to galactic nuclei, despite tens of years of observational, analytical, and numerical studies, is still a matter of debate due to, e.g., the presence of external pressure and the tidal field among other factors \citep{2016ApJ...819..138C}. Therefore, we examine the aspect of the BLR cloud formation and its stability during the expected flight time within the FRADO model below in some detail.

\subsection{BLR Cloud Thermal Stability}\label{sec:BLR_cloud_stability}

The inner gravitational radii where the BLR clouds originate are characterized by a multiphase medium. The low-ionization BLR clouds are dense structures with $n_{\rm c}\sim 10^{12}\,{\rm cm^{-3}}$, whereas the typical disk densities are $n_{\rm disk}\simeq 10^{14}-10^{15}\,{\rm cm^{-3}}$ (see Fig.~\ref{fig:densities}), where we considered a standard thin disk with $\dot m = 1$, $M_{\rm BH}=10^8\,M_{\odot}$, and the BLR range of $r=5\,000-20\,000\,r_{\rm g}$ according to the calculations by \citet{naddaf2021}. The BLR clouds are lifted from the accretion disk into the hot intercloud medium, where we assume they are approximately in pressure equilibrium. 

The BLR gas temperature is $T_{\rm c}\simeq 10^4\,{\rm K}$ based on the heating/cooling balance due to atomic transitions in the partially ionized plasma \citep{2006LNP...693...77P,2019OAst...28..200C}. The intercloud medium is kept close to the Compton limit of a few $\sim 10^7\,{\rm K}$ due to heating/cooling by the Compton scattering \citep{1981ApJ...249..422K}. The irradiation by the intense hard central source is expected to lead to the development of the thermal instability and the two phases above the disk plane coexist with the number density ratio of $n_{\rm c}/n_{\rm h}\sim 10^3$, where $n_{\rm h}\sim n_{\rm c}\,T_{\rm c}/T_{\rm h}\sim 10^9\,{\rm cm^{-3}}$ is the number density of the hot medium inferred from the assumed pressure equilibrium.

First, we consider the motion of a BLR cloud through a hot, stationary plasma. In that case, the ablation of the cloud likely takes place due to its interception by the hot plasma with a certain velocity difference. The mass-loss rate of the cloud then can be expressed as \citep{2011ApJ...739...30K,2012ApJ...750...58B},
\begin{equation}
    \frac{\mathrm{d}M_{\rm c}}{\mathrm{d}t}\simeq -q_{\rm abl}\,\pi\,R_{\rm c}^2\,  
    m_p\,n_{\rm h}\,v_{\rm c}\,,
    \label{eq_ablation_massloss}
\end{equation}
where $v_{\rm c}$ is the cloud velocity with the dominant Keplerian component at the launch radius and $q_{\rm abl}$ is the ablation coefficient. Assuming that $n_{\rm c}$ is constant, the cloud radius evolution is 
\begin{equation}
    \frac{\mathrm{d}R_{\rm c}}{\mathrm{d}t}=-\frac{q_{\rm abl}}{4}v_{\rm c}\frac{\rho_{\rm h}}{\rho_{\rm c}} 
    \label{eq_radius_ablation}
\end{equation}
\noindent and the ablation timescale is
\begin{align}
    \tau_{\rm abl} &= \frac{4R_{\rm c}}{q_{\rm abl}v_{\rm c}}\frac{\rho_{\rm c}}{\rho_{\rm h}}\,\notag\\
    &\sim 106\,\left(\frac{R_{\rm c}}{10^{12}\,{\rm cm}} \right)\left(\frac{v_{\rm c}}{3000\,{\rm km\,s^{-1}}} \right)^{-1}\left(\frac{\rho_{\rm c}/\rho_{\rm h}}{10^3}\right)\,{\rm yr}\,,\label{eq_ablation_timescale}
\end{align}
where we fixed $q_{\rm abl} =  0.004$ \citep{2011ApJ...739...30K}.

A small BLR cloud surrounded by the hot medium also loses mass and evaporates due to the thermal conduction. The thermal conduction can be treated in a classical diffusion approximation since the saturation parameter \citep{1977ApJ...211..135C}
\begin{align}
    \sigma_0 &= 1.84\frac{\lambda_{\rm f}}{R_{\rm c}}\,\notag\\
       &\simeq 10^{-3} \left(\frac{T_{\rm h}}{10^7\,K}\right)^2 \left(\frac{R_{\rm c}}{10^{12}\,{\rm cm}}\right)^{-1} \left(\frac{n_{\rm h}}{10^9\,{\rm cm^{-3}}}\right)^{-1}\,,\label{eq_saturation_param1}
\end{align}
\noindent where $\lambda_{\rm f}$ is the  electron mean-free path, is smaller than one. Using the classical diffusion approximation below the saturation level, the evaporation timescale can be estimated as \citep{1977ApJ...211..135C}
\begin{align}
    \tau_{\rm evap} &\sim 110 \left(\frac{n_{\rm c}}{10^{12}\,{\rm cm^{-3}}}\right)\left(\frac{R_{\rm c}}{10^{12}\,{\rm cm}} \right)^2\times \notag\\
    &\times\left(\frac{T_{\rm h}}{10^7\,{\rm K}} \right)^{-\frac{5}{2}} \left(\frac{\log{\Lambda}}{30} \right)\,{\rm yr}\,,
    \label{eq_tau_evaporation}
\end{align}
where $\log{\Lambda}$ is the Coulomb logarithm.

Due to the velocity shear $v_{\rm shear}$ between the BLR clouds and the hot medium, or the shock created in a supersonic motion, which can be estimated to be equal to the initial launch Keplerian velocity $v_{\rm c}$, the clumps are subject to Kelvin-Helmholtz (KH) instabilities with the characteristic size of $\lambda_{\rm KH}$. Given the density ratio between the medium and the clouds, $r_{\rho}=n_{\rm h}/n_{\rm c}\sim 10^{-3}$, the characteristic timescale $\tau_{\rm KH}$ for the growth of instabilities of the size $\lambda_{\rm KH}\sim R_{\rm c}$ can be calculated as \citep[see e.g.][]{2015MNRAS.449....2M,2020ApJ...897...28P}
\begin{align}
    \tau_{\rm KH}&=\frac{\lambda_{\rm KH}}{v_{\rm shear}}\frac{1+r_{\rho}}{\sqrt{r_{\rho}}}\,\notag\\
    &=3.3\times 10^{-3} \left(\frac{\lambda_{\rm KH}}{10^{12}\,{\rm cm}} \right)\left(\frac{v_{\rm shear}}{3000\,{\rm km\,s^{-1}}} \right)^{-1}\times \notag\\
    &\times \frac{1+r_{\rho}}{\sqrt{r_{\rho}}}\,{\rm yr}\sim 1.2\,\mathrm{days}.
    \label{eq_tau_KH}
\end{align} 
The dynamical timescale, on which any perturbation inside the cloud propagates, is 
\begin{equation}
    \tau_{\rm dyn} =\frac{R_{\rm c}}{c_{\rm c}}=0.025\,\left(\frac{R_{\rm c}}{10^{12}\,{\rm cm}} \right)\left(\frac{\mu}{0.5}\right)^{\frac{1}{2}}\left(\frac{T_{\rm c}}{10^{4}\,{\rm K}} \right)^{-\frac{1}{2}}\,{\rm yr}\,\label{eq_dyn_timescale}
\end{equation}
where $c_{\rm c}$ denotes the sound speed inside the BLR cloud and we adopted the mean molecular weight of $\mu=0.5$ corresponding to the ionized hydrogen plasma.

\begin{figure}[tbh!]
    \centering
    \includegraphics[trim={0.25cm 0.25cm 0.25cm 0.25cm},clip,width=\columnwidth]{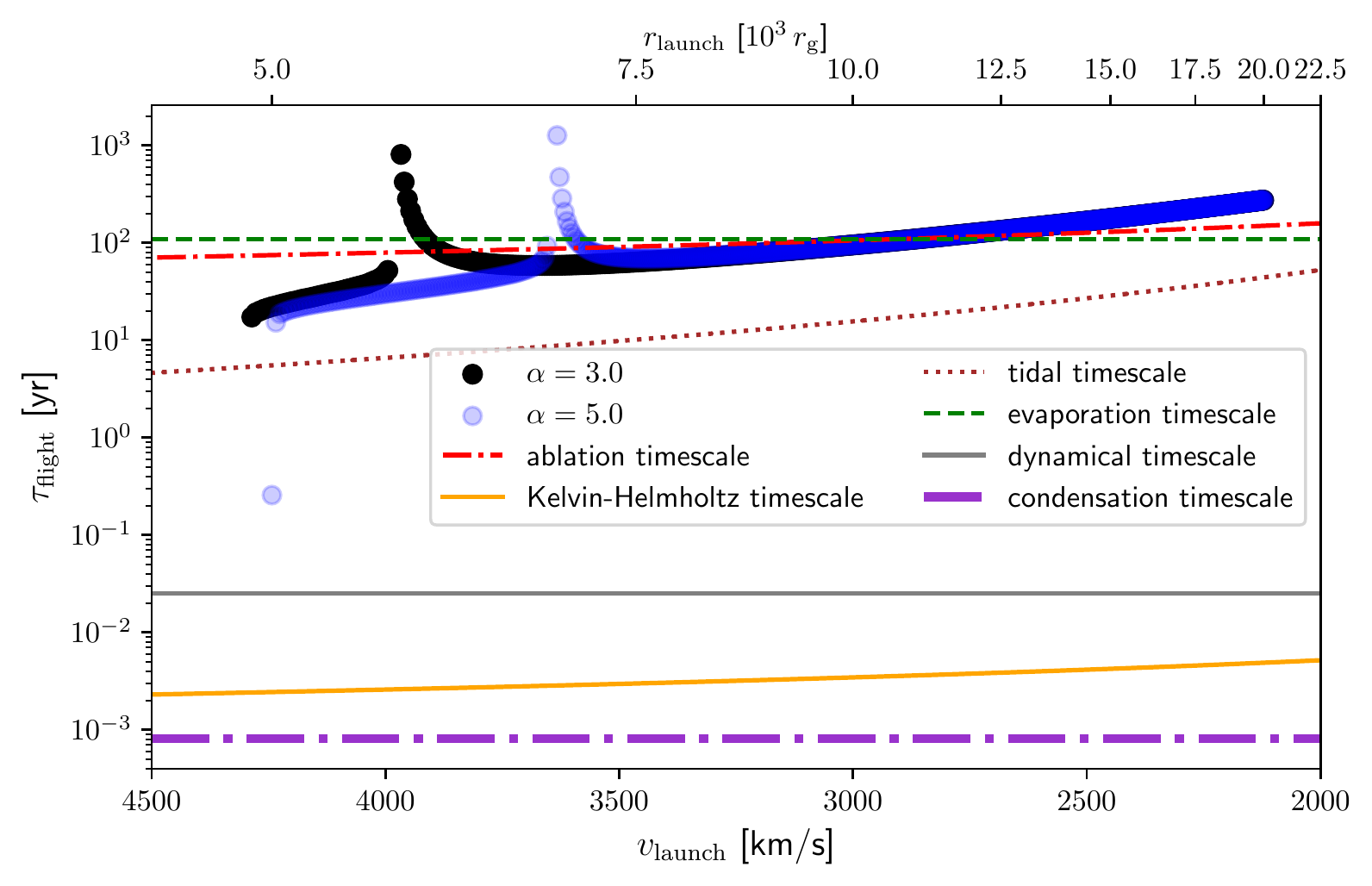}
    \caption{Flight time of BLR clouds as a function of the launch Keplerian velocity for $M_{\rm BH}=10^8\,M_{\odot}$, $\dot{m}=1$, and for two shielding parameter values ($\alpha=3$ and $5$). The ablation, evaporation, Kelvin-Helmholtz, tidal, dynamical, and condensation timescales are shown by lines according to the legend. The top $x$-axis is the launch radius in gravitational radii.}
    \label{fig_tau_vel}
\end{figure}

In Fig.~\ref{fig_tau_vel}, we plot the cloud flight time $t_{\rm flight}$ as a function of the launch Keplerian velocity that is a dominant velocity component during the motion. For these exemplary calculations, we consider $M_{\rm BH} = 10^8\,M_{\odot}$, $\dot m=1$, and two different shielding parameter values $\alpha=3$ (black points) and $5$ (blue points) \citep{naddaf2021}. For comparison, we also plot $\tau_{\rm abl}$, $\tau_{\rm evap}$, $\tau_{\rm KH}$, and $\tau_{\rm dyn}$ given in Eqs.~\eqref{eq_ablation_timescale}, \eqref{eq_tau_evaporation}, \eqref{eq_tau_KH}, and \eqref{eq_dyn_timescale}, respectively. We can see that $\tau_{\rm abl}\sim \tau_{\rm evap}$ and in some cases they are shorter than $t_{\rm flight}$. On the other hand, the Kelvin-Helmholtz timescale is at least four orders of magnitude shorter than $t_{\rm flight}$. This implies that upon falling back onto the accretion disk, the clouds are expected to be diminished, deformed and in some cases, they can be completely ablated and evaporated before their trajectories are completed. In case of the existence of an entangled magnetic field inside the clouds, it can effectively inhibit the growth of Kelvin-Helmholtz instabilities and prolong the cloud lifetime \citep{2015MNRAS.449....2M}. However, the ablation as well as the evaporation are not expected to be significantly mitigated by the internal magnetic field. The BLR clouds can further be affected significantly by the tidal forces that stem from the presence of the SMBH, which we analyze in Subsection~\ref{subsec_tidal}. The destructive effect of the ablation and the evaporation is expected to be reversed by the condensation of the hot intercloud medium on the ``seed'' clouds as we show in Subsection~\ref{ref_cloud_condensation}.

\subsection{Tidal Deformation of Clouds}
\label{subsec_tidal}

The shape and the stability of BLR clouds is also affected by the tidal forces in the vicinity of the SMBH. BLR clouds are about ten orders of magnitude less massive than the critical Bonnor-Ebert mass for the collapse of an isothermal cloud, $M_{\rm BE}\sim c_{\rm BE}v_{\rm T}^4/(P_{\rm ext}^{1/2}G^{3/2})$, where $c_{\rm BE}\simeq 1.18$, $v_{\rm T}$ is the thermal velocity inside an isothermal BLR cloud, and $P_{\rm ext}$ is the external pressure, including the ram pressure. Assuming a BLR cloud of $T_{\rm c}\sim 10^4\,{\rm K}$, which is moving at $v_{\rm c}\sim 3000\,{\rm km\,s^{-1}}$ in a hot medium of $n_{\rm h}\sim 10^9\,{\rm cm^{-3}}$ and $T_{\rm h}\sim 10^7\,{\rm K}$, we obtain $M_{\rm c} \ll M_{\rm BE}\sim 11\,M_{\odot}$.

The BLR clouds are likely not tidally stable along their trajectory across a range of distances (see Fig.~\ref{fig_tau_vel}). We obtain the Roche critical density taking only into account the radial motion \citep{1989IAUS..136..543P,1998MNRAS.294...35S,2020ApJ...897...28P},
\begin{align}
    n_{\rm Roche} &\sim \frac{3M_{\rm BH}}{2\pi \mu m_{\rm p}r^3}\,\notag
              =\frac{3c^6}{2 \pi G^3 \mu m_{\rm p} (r/r_{\rm g})^3 M_{\rm BH}^2}\,\notag\\
              &\sim 3.5\times 10^{13}\left(\frac{r}{10^4\,r_{\rm g}} \right)^{-3} \left(\frac{M_{\rm BH}}{10^8\,M_{\odot}} \right)^{-2}\!{\rm cm^{-3}}\!.\label{eq_Roche}
\end{align}
Therefore, BLR clouds with $n_{\rm c}\simeq 10^{12}\,{\rm cm^{-3}}<n_{\rm Roche}$ are tidally unstable. However, at distances larger by a factor of two than $10^4\,r_{\rm g}$, tidal stability could be reached in case the density is not reduced; hence, the stability with respect to tidal effects and the associated cloud distortion depend strongly on the distance.

Another indicator of the importance of tidal effects is the tidal radius $r_{\rm tidal}$, which depends on the mass of the cloud as well as linearly on its size as follows \citep{1975Natur.254..295H,2020arXiv200512528R}
\begin{align}
    r_{\rm tidal} &= R_{\rm c} \left(\frac{M_{\rm BH}}{M_{\rm c}} \right)^\frac{1}{3}=\left(\frac{4}{3}\pi \mu  m_{\rm p}\right)^{-\frac{1}{3}}\left(\frac{M_{\rm BH}}{n_{\rm c}}\right)^\frac{1}{3}\,\notag\\
    & \simeq 2.6\times 10^4 \left(\frac{M_{\rm BH}}{10^8\,M_{\odot}} \right)^\frac{1}{3} \left(\frac{n_{\rm c}}{10^{12}\,{\rm cm^{-3}}} \right)^\frac{1}{3}\,r_{\rm g}\,,
    \label{eq_tidal_radius}
\end{align}
which indicates that BLR clouds orbiting at distances $r\lesssim r_{\rm tidal}$ are prone to tidal effects, specifically tidal stretching as well as compression. The acceleration acting on a BLR cloud of size $R_{\rm c}$ (in the direction of the SMBH) due to the tidal field can be expressed as
\begin{equation}
    |a_{\rm tidal}|=2R_{\rm c}\frac{GM_{\rm BH}}{r^3}\,,
    \label{eq_tidal_acceleration}
\end{equation}
which implies that the timescale on which the tidal deformation of the size $R_{\rm c}$ develops is
\begin{align}
    \tau_{\rm tidal} &= \sqrt{\frac{2R_{\rm c}}{|a_{\rm tidal}|}}\notag\\
    &=\frac{r^{3/2}}{\sqrt{GM_{\rm BH}}}=\left(\frac{r}{r_{\rm g}}\right)^\frac{3}{2}\frac{GM_{\rm BH}}{c^3}\notag\\
    &\simeq 15.6 \left(\frac{r}{10^4\,r_{\rm g}} \right)^\frac{3}{2} \left(\frac{M_{\rm BH}}{10^8\,M_{\odot}} \right)\,{\rm yr}\,.\label{eq_tidal_timescale}
\end{align}
The BLR clouds are considered to be non-selfgravitating. Each cloud becomes significantly stretched by $R_{\rm c}$ on a timescale shorter than or comparable to the orbital timescale $P_{\rm orb}$. The characteristic tidal timescale expressed by Eq.~\eqref{eq_tidal_timescale} is depicted in Fig.~\ref{fig_tau_vel}, where $t_{\rm flight}$ is typically longer than $\tau_{\rm tidal}$. For the shielding models with $\alpha=3$ and 5, the mean ratio of the flight time to the orbital time is $\tau_{\rm flight}/P_{\rm orb}=1.06 \pm 0.80$ and $1.03 \pm 0.88$, respectively. Hence, the mean flight time is of the order of the orbital timescale. From the formation of the cloud till its fallback onto the disk, the tidal field can induce stretching by a factor of at least two.

\begin{figure*}[tbh!]
    \centering
    \includegraphics[trim={0.24cm 0.24cm 0.24cm 0.24cm},clip, width=0.32\textwidth]{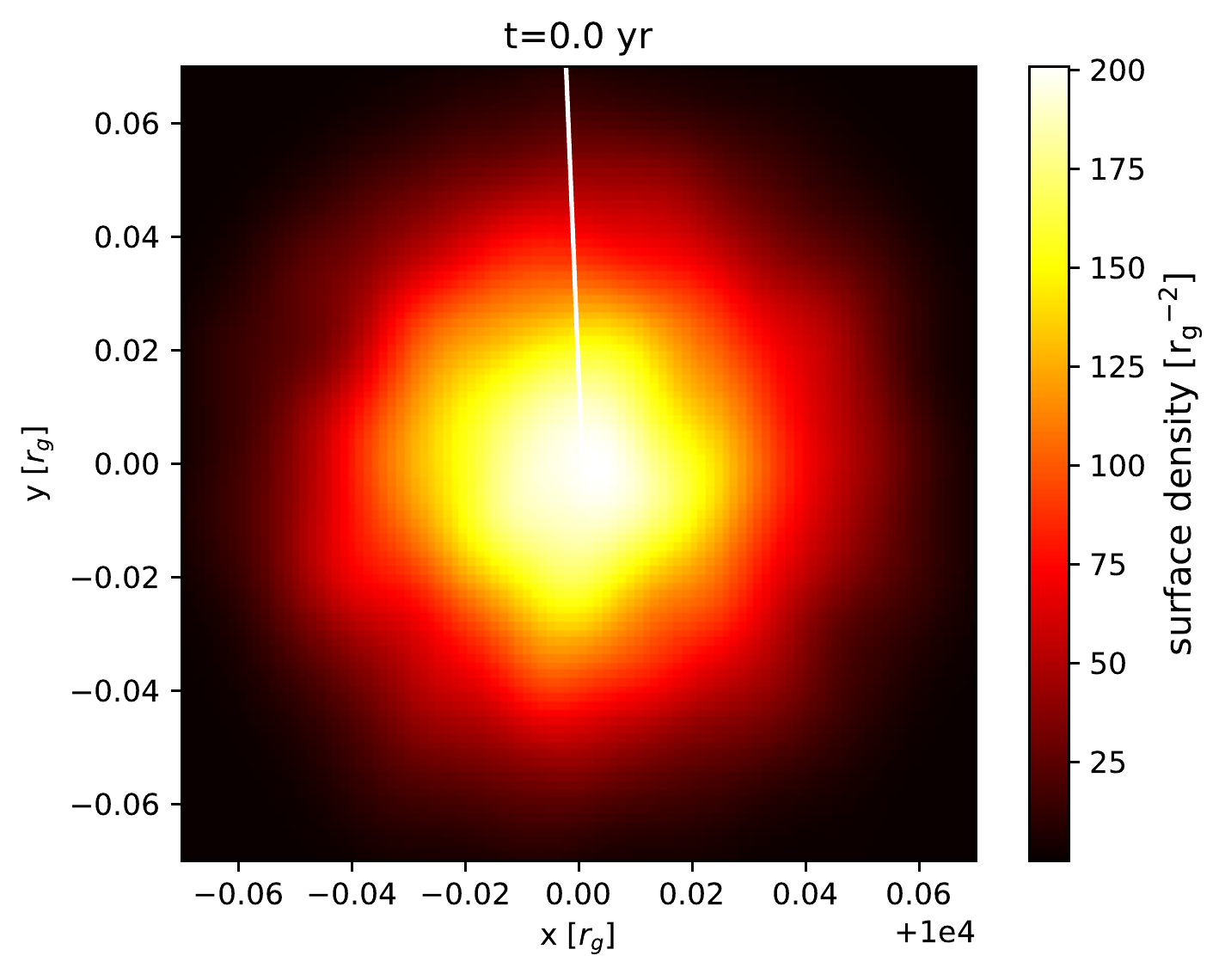}
    \includegraphics[trim={0.24cm 0.24cm 0.24cm 0.24cm},clip,width=0.32\textwidth]{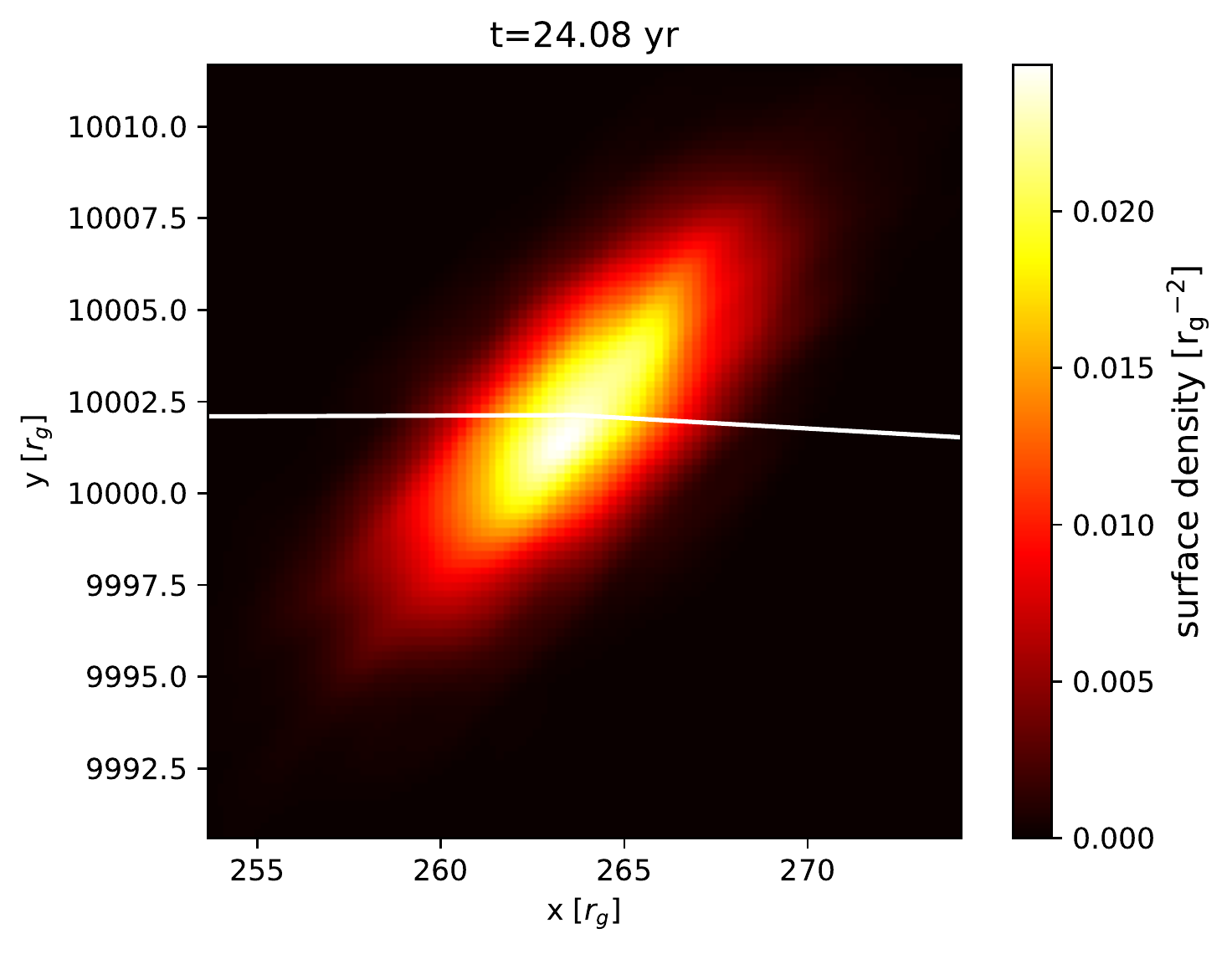}
    \includegraphics[trim={0.24cm 0.24cm 0.24cm 0.24cm},clip,width=0.2\textwidth]{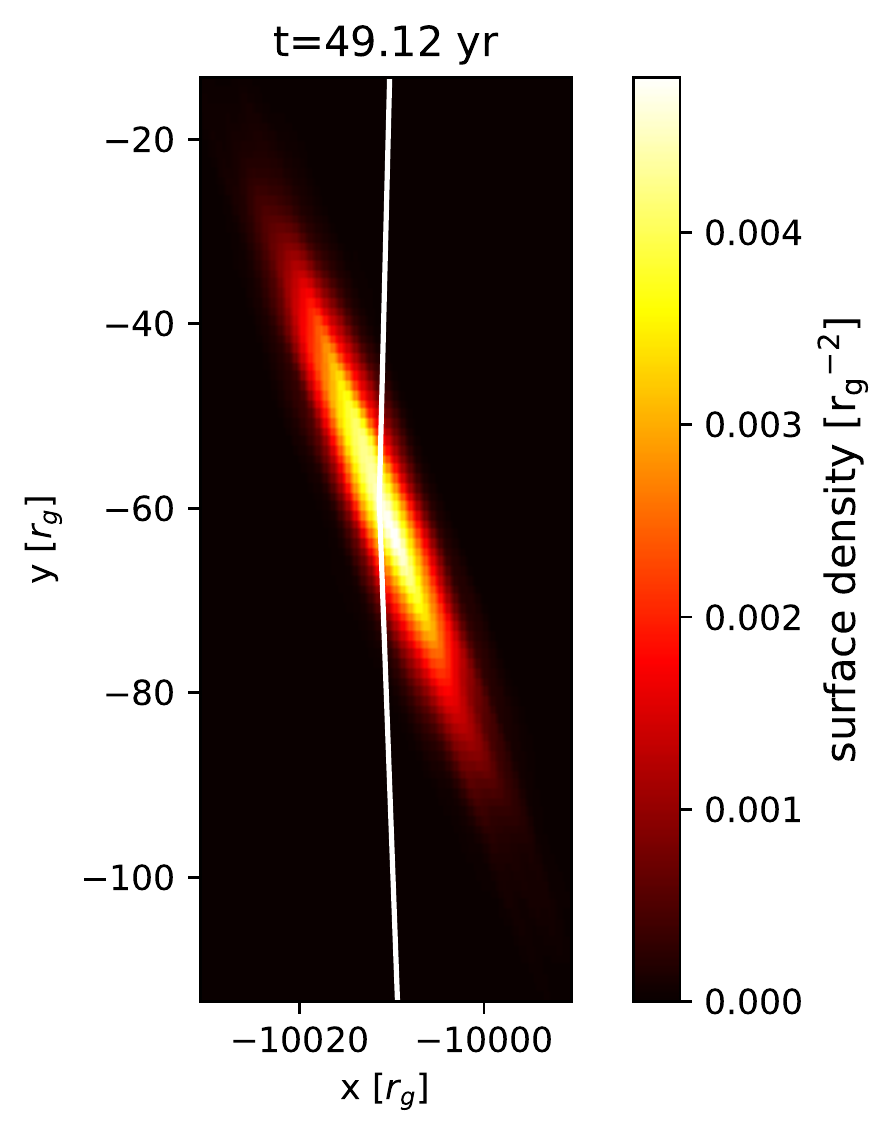}
    \caption{The evolution of a simulated BLR cloud of free particles with a small velocity dispersion of $1\,{\rm km\,s^{-1}}$ in the orbital plane. The cloud orbits the SMBH of $10^8\,M_{\odot}$ at $10^{4}\,r_{\rm g}$. We show the colour-coded distribution of the particle density at  $t=0$, $24.08$, and $49.12\,{\rm yr}$, which corresponds to the initial time, the quarter and the half of the orbital period, respectively. The $x$ and $y$ axes scaling changes according to the size of the tidally elongated cloud. The colour axis scale also changes according to the peak surface density, which drops as the cloud is being tidally stretched due to the central potential.}
    \label{fig_BLR_cloud_evolution_small_disp}
\end{figure*}

\begin{figure*}[tbh!]
    \centering
    \includegraphics[width=0.32\textwidth]{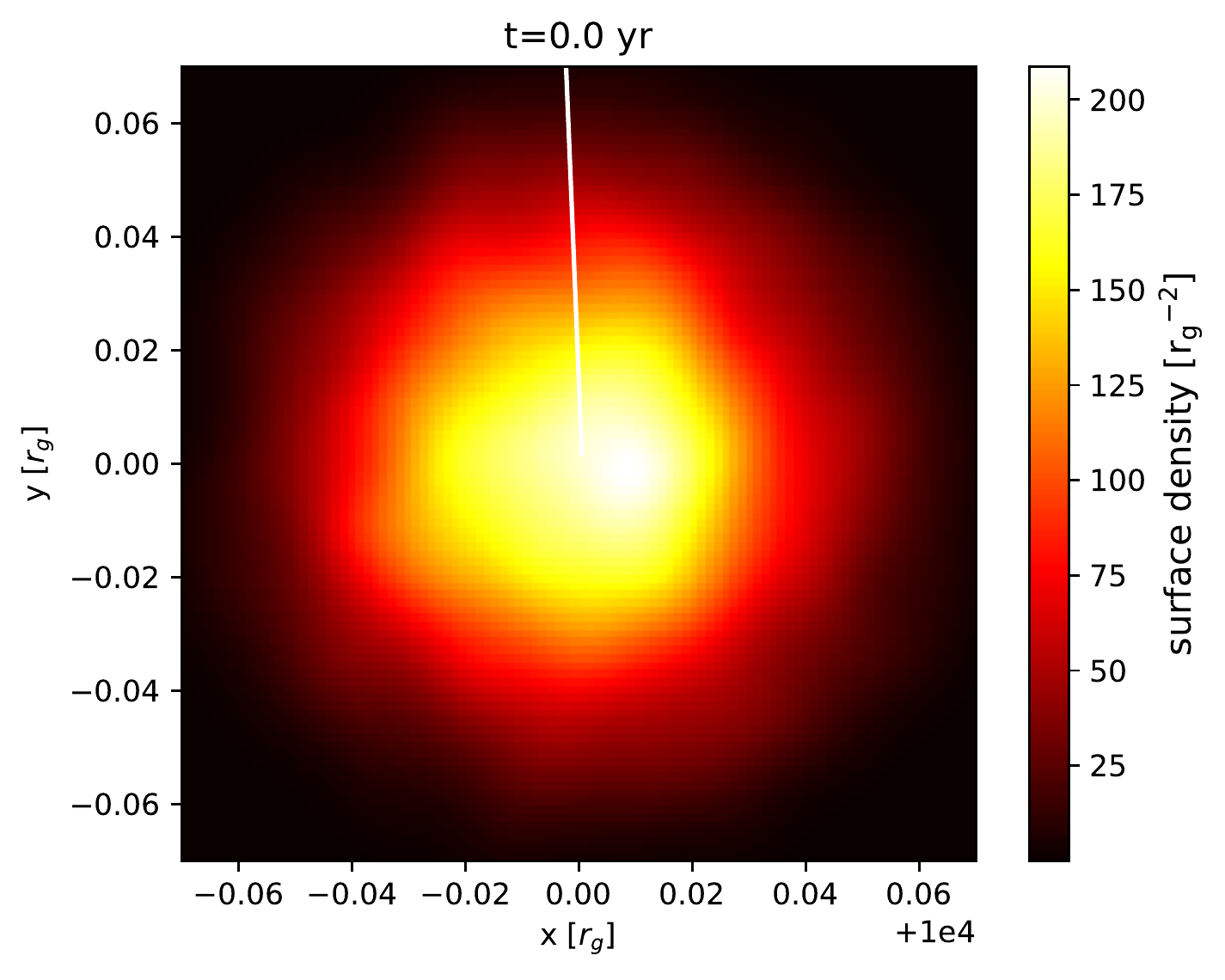}
    \includegraphics[width=0.32\textwidth]{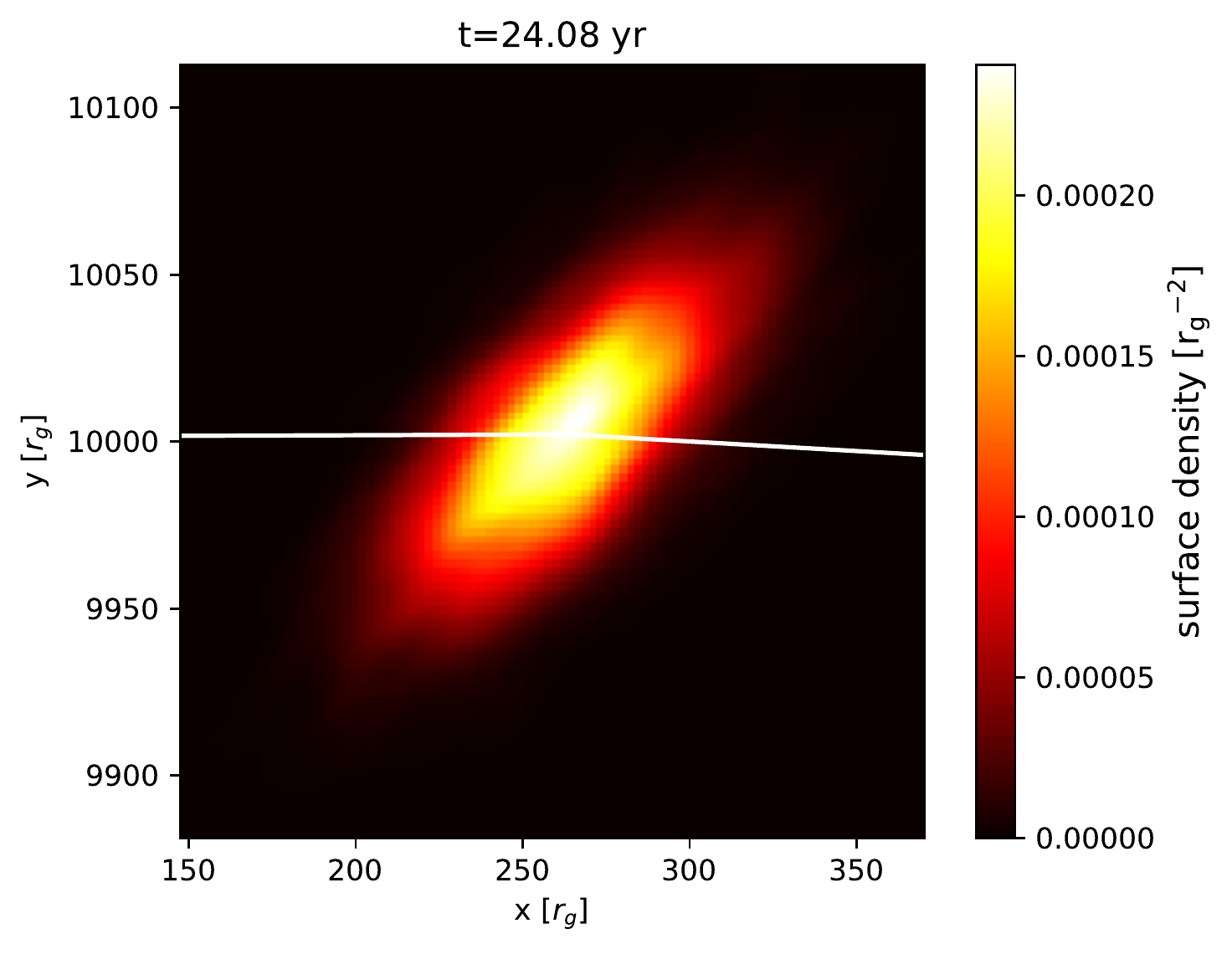}
    \includegraphics[width=0.2\textwidth]{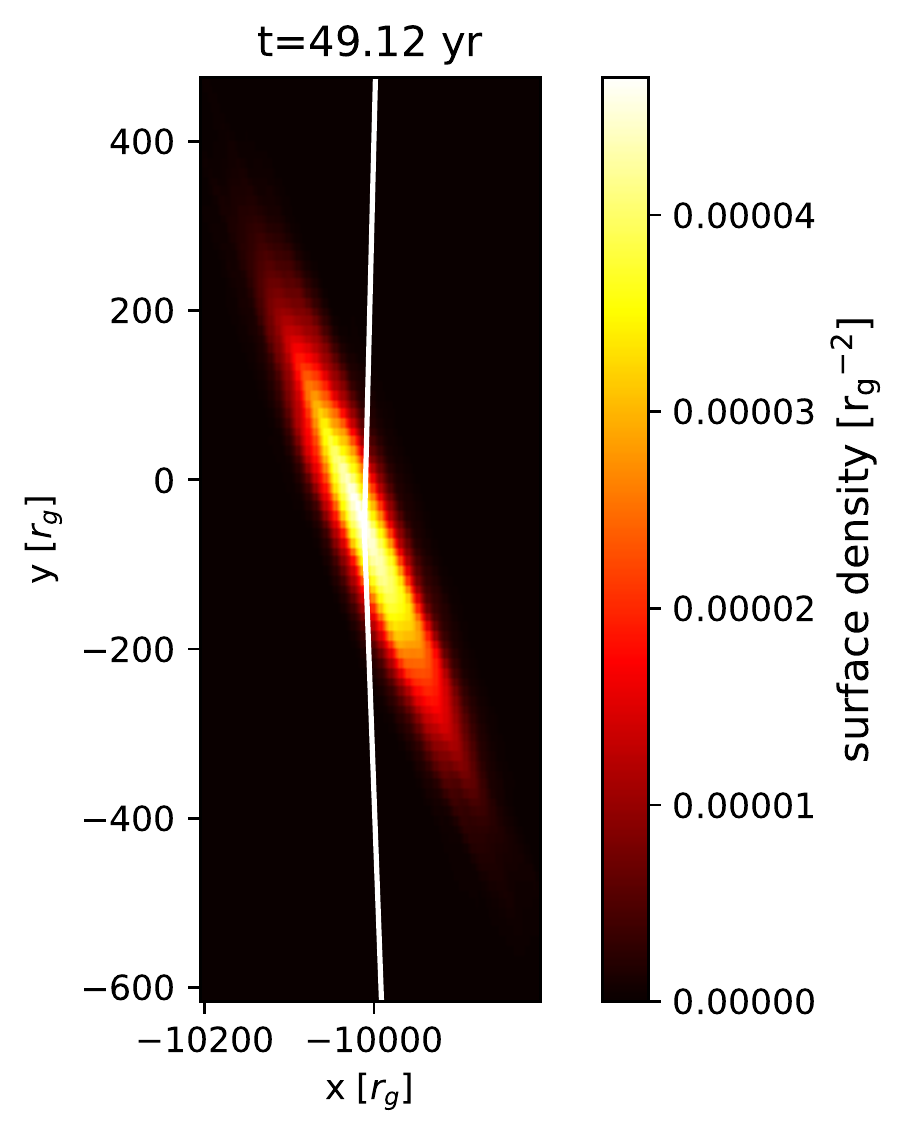}
    \caption{The same as in Fig.~\ref{fig_BLR_cloud_evolution_small_disp}, but for the velocity dispersion of $10\,{\rm km\,s^{-1}}$. Note the difference in the colour scale with respect to Fig.~\ref{fig_BLR_cloud_evolution_small_disp}.}
    \label{fig_BLR_cloud_evolution_large_disp}
\end{figure*}

\begin{figure*}[tbh!]
    \centering
    \includegraphics[width=0.32\textwidth]{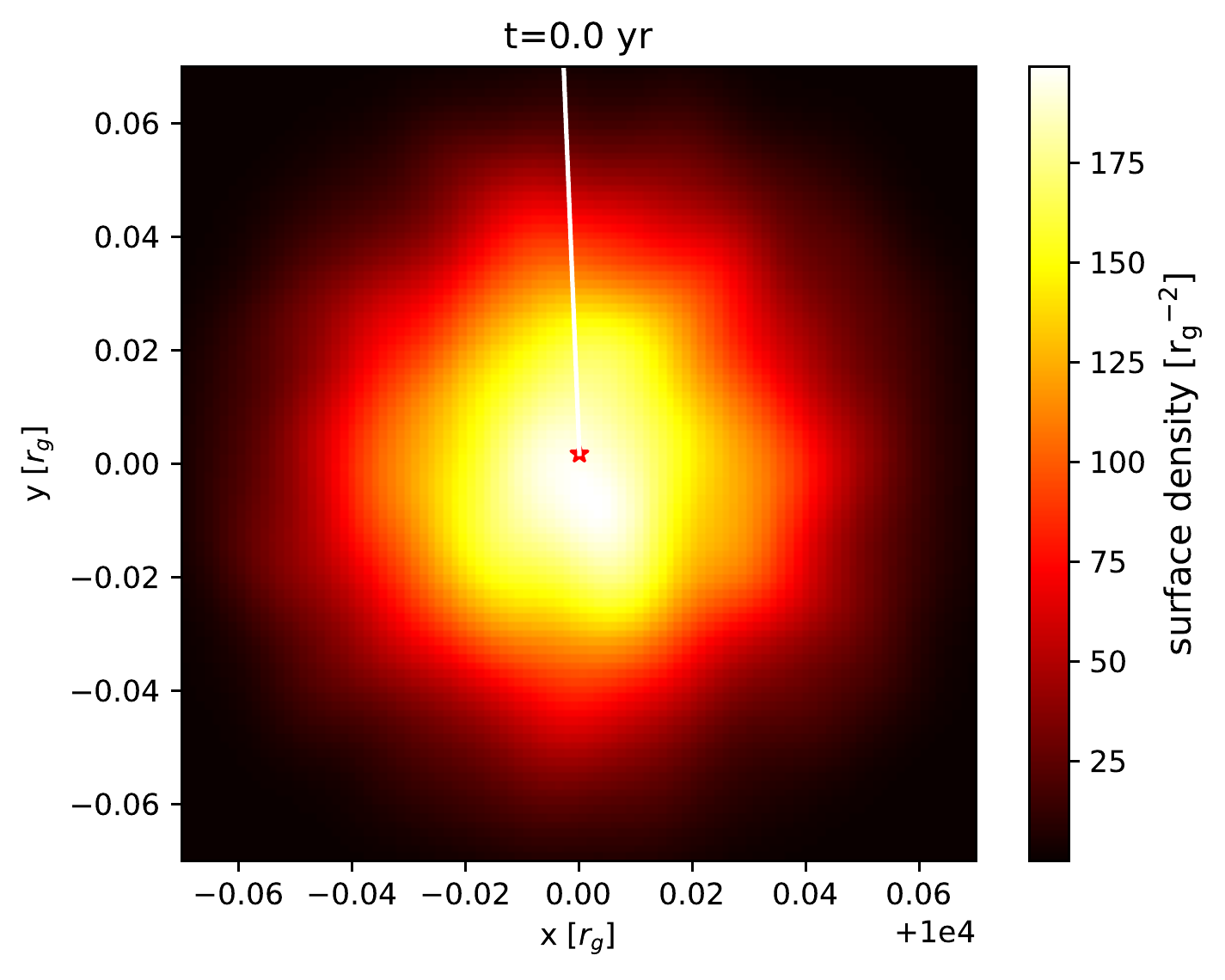}
    \includegraphics[width=0.32\textwidth]{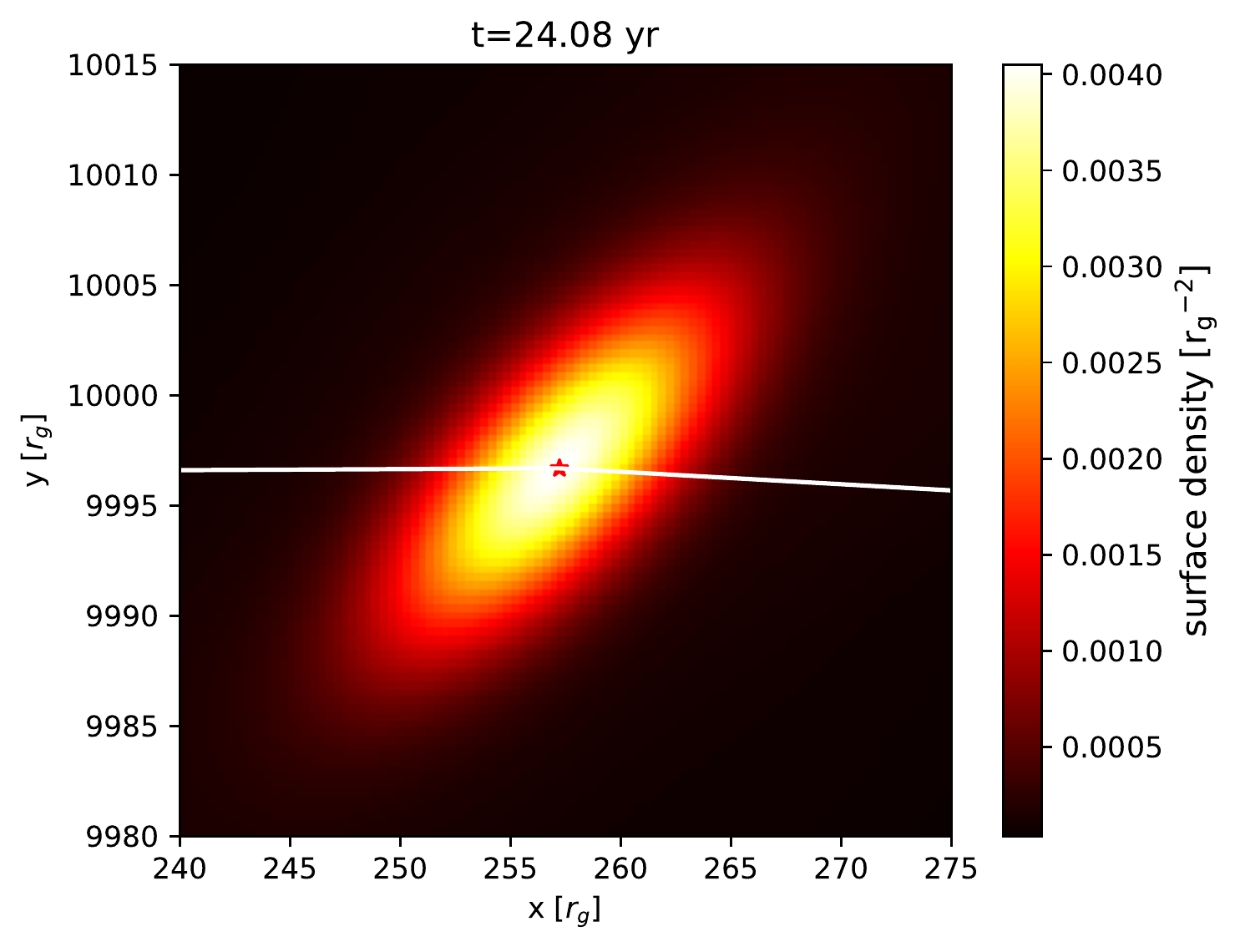}
    \includegraphics[width=0.21\textwidth]{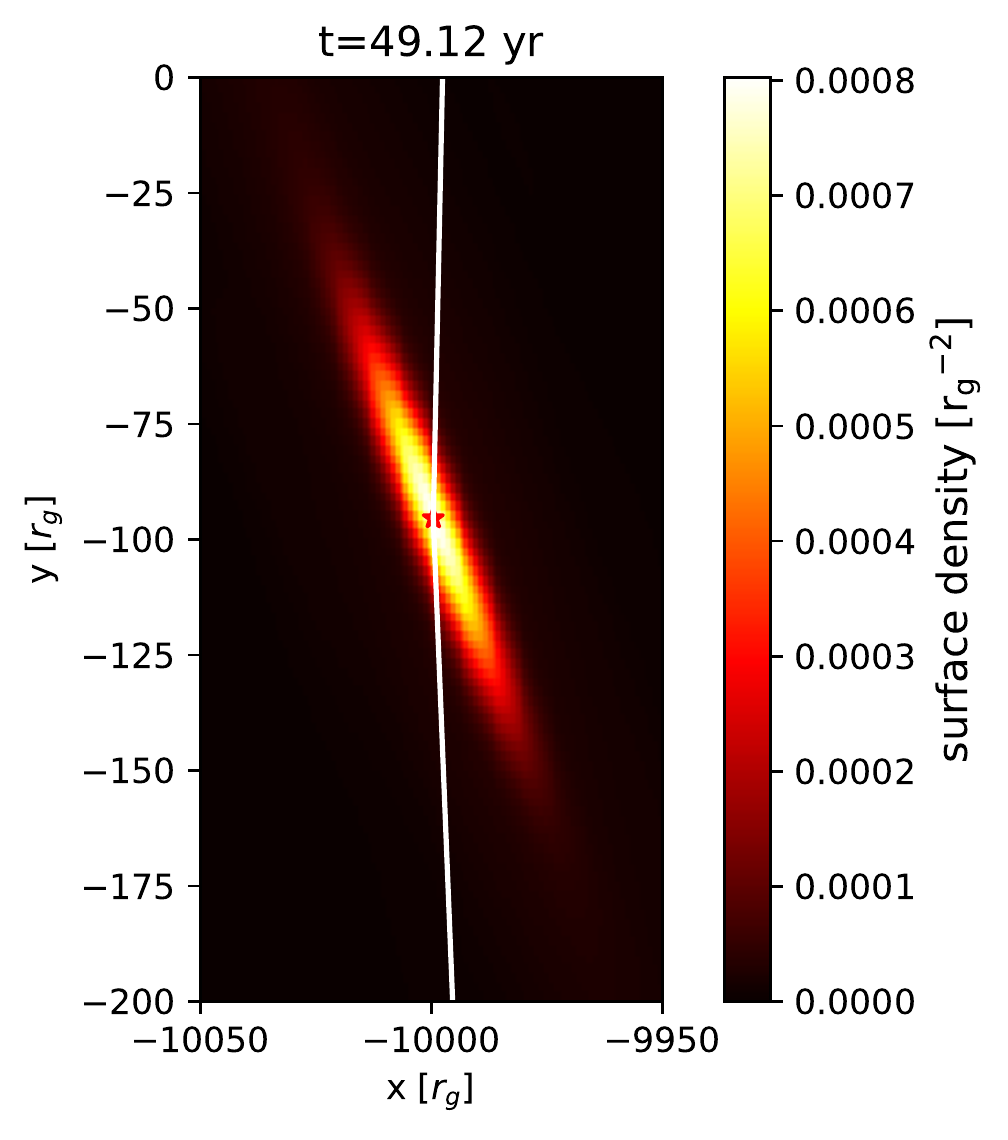}
    \caption{The same as in Fig.~\ref{fig_BLR_cloud_evolution_large_disp}, but including a central core of mass $10^{-3}\,M_{\odot}$. Note the difference in the colour scale with respect to Fig.~\ref{fig_BLR_cloud_evolution_large_disp}.}
    \label{fig_BLR_cloud_evolution_core}
\end{figure*}

We examine the tidal effects by a test-particle simulation, where we initially distribute particles in a spherical cloud with a Gaussian density distribution of Full Width at Half Maximum (FWHM) size of $10^{12}\,{\rm cm}$, set at the expected cloud size (the corresponding width of the Gaussian is $4.3 \times 10^{11}$ cm), and a small velocity dispersion of $1\,{\rm km\,s^{-1}}$. The cloud as a whole orbits around the SMBH of $10^8\,M_{\odot}$ at a distance of $10^4\,r_{\rm g}$ (with an orbital period of $\sim 98$ years). In Fig.~\ref{fig_BLR_cloud_evolution_small_disp} we show several snapshots of the cloud shape and its surface density distribution at $t=0$, $24.08$ (a quarter of the orbital period), and $49.12\,{\rm yr}$ (half of the orbital period). We see that a BLR cloud without any additional support would become more and more elongated, with a progressively decreasing surface density. Initially the cloud is spherical with a radius of $R_{\rm c}\sim 0.07\,r_{\rm g}$, at the quarter of the period it is elongated with the proportions $\sim (9\times 3 \times 5)\,r_{\rm g}$ (radii across the long, short axis and the vertical direction), and at the half of the period, it gets prolonged to $\sim (35\times 4\times 0.09)\,r_{\rm g}$.

The effect of the surface density decrease is more profound for a larger particle velocity dispersion of $10\,{\rm km\,s^{-1}}$, which corresponds to the sound speed inside a BLR cloud of $T_{\rm c}\sim 10^4\,{\rm K}$. At the same epochs, the aspect ratio is comparable to a smaller dispersion, but the spatial extension is larger and hence the surface density drop is larger by two orders of magnitude, see Fig.~\ref{fig_BLR_cloud_evolution_large_disp}. Placing a central core of mass $m_{\rm core}=10^{-3}\,M_{\odot}$ at the center of the cloud makes it more compact for the same velocity dispersion, see Fig.~\ref{fig_BLR_cloud_evolution_core}. However, the tidal distortion still proceeds on the spatial scale comparable to a dispersion of $1\,{\rm km\,s^{-1}}$. The tidal (Hill) radius for a core of $m_{\rm core}=10^{-3}\,M_{\odot}$ is $r_{\rm Hill}=a[m_{\rm core}/(3M_{\rm BH})]^{1/3}\sim 2.2\times 10^{13}\,{\rm cm}$, which encompasses the whole BLR cloud. 

Due to the tidal field, the BLR cloud will effectively stretch, lose mass to the surroundings, and its surface density will drop. In this sense, the mass loss and the cloud dissolution due to tidal stretching accompanies the mass loss due to the ablation and the evaporation (see Eqs.~\eqref{eq_ablation_timescale} and \eqref{eq_tau_evaporation}, respectively). Nevertheless, these processes appear to be swiftly compensated by condensation within the BLR region. We discuss this possibility in the following section.

\subsection{Cloud Condensation}
\label{ref_cloud_condensation}

Clouds are not only a subject of destruction: they are expected to gain mass via the condensation from the hot surroundings in the thermal instability regime. Whether the condensation or the evaporation prevail is given by the global saturation parameter $\sigma_0$, which expresses the ratio of the electron mean free path to the cloud radius $R_{\rm c}$ \citep{1977ApJ...211..135C}
\begin{equation}
    \sigma_0=\frac{0.08\kappa_{\rm h}T_{\rm h}}{\Phi_{\rm s}\rho_{\rm h}c_{\rm h}^3 R_{\rm c}}\,,
    \label{eq_saturation_param2}
\end{equation}
where $\kappa_{\rm h}$ is the thermal conductivity of the hot medium, $\Phi_{\rm s}$ is a factor of the order of unity, and $c_{\rm h}$ is the sound speed in the hot medium; see also Eq.~\eqref{eq_saturation_param1} for comparison. For the conductivity, we use \citep{1962pfig.book.....S}
\begin{equation}
    \kappa_{\rm h}=\frac{1.84\times 10^{-5}T_{\rm h}^{5/2}}{\ln{\Psi}}\,{\rm erg\,s^{-1}\,K^{-1}\,cm^{-1}}\,,
    \label{eq_conductivity}
\end{equation}
where $\ln{\Psi}=29.7+\ln{(T_{6,e}/\sqrt{n_{\rm e}})}$ with $T_{6,e}$ being the electron temperature in $10^6\,{\rm K}$ and $n_{\rm e}$ the electron number density. Using Eq.~\eqref{eq_conductivity}, we obtain $\sigma_0\simeq 4\times 10^{-3}/\Phi_{\rm s}<0.027/\Phi_{\rm s}$, which implies that the material condenses onto the BLR cloud since the radiative cooling losses overcome the conductive heat input \citep{2007A&A...472..141V}.

Equivalently, the condition for the survival of clouds can be expressed in terms of the Field length $\lambda_{\rm F}$ \citep{1965ApJ...142..531F,1970ApJ...160..659D,1990ApJ...358..375B}
\begin{equation}
    \lambda_{\rm F}=\sqrt{\frac{\kappa_{\rm h}T_{\rm h}}{n^2_{\rm h}\Lambda_{\rm h}}}\,,
    \label{eq_Field_length}
\end{equation}
\noindent where $\Lambda_{\rm h}(T_{\rm h})\sim 10^{-23}\,{\rm erg\,s^{-1}\,cm^{3}}$ is the cooling function value for the hot medium temperature of $T_{\rm h}\sim 10^7\,{\rm K}$. For the conductivity in the hot medium, we obtain $\kappa_{\rm h}\sim 2.7 \times 10^{11}\,{\rm erg\,s^{-1}\,K^{-1}\,cm^{-1}}$, using Eq.~\eqref{eq_conductivity}. Finally, according to Eq.~\eqref{eq_Field_length}, the Field length is $\lambda_{\rm F}\sim 5.2\times 10^{11}\,{\rm cm}$. Since the cloud radius is $R_{\rm c}\sim 10^{12}\,{\rm cm}\gtrsim \lambda_{\rm F}$, the hot intercloud medium will rather condense on the cloud surface since the radiative cooling prevails over the thermal conduction \citep{2014MNRAS.445.4385R,2017MNRAS.464.2090R}. The cloud size will tend to be progressively larger than $\lambda_{\rm F}$ during the flight time due to the tidal stretching, which will provide a greater area for the condensation to take place \citep[cp. also][where filamentary multiphase gaseous structure have been explored in a different context of black hole accretion]{2012MNRAS.424..728B}. \citet{McCourt2018} argued that the thermal instability will always lead to the creation of the smallest possible clouds (with sizes much smaller than the Field length) in the form of a mist, but this may be related to their simplified treatment of the electron conduction and the neglected role of the magnetic field.

\begin{figure}[h!]
    \centering
    \includegraphics[width=0.5\textwidth]{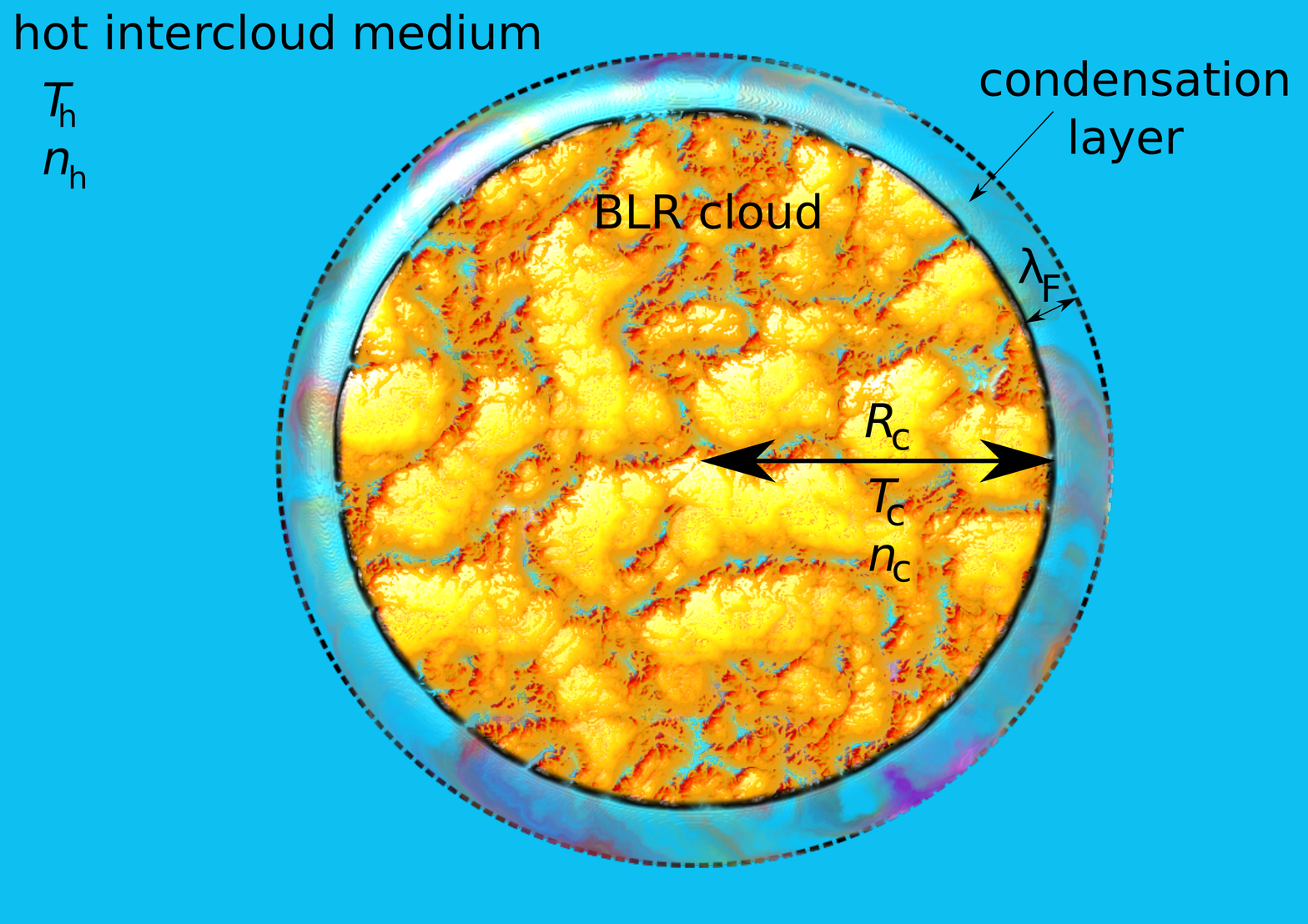}
    \caption{A cross-sectional view of the spherical BLR-cloud model that is characterized by $R_{\rm c}$, $T_{\rm c}$, and $n_{\rm c}$ and that is embedded within the hot intercloud medium with $T_{\rm h}$ and $n_{\rm h}$. The condensation layer around the cloud has the length-scale of $\lambda_{\rm F}$, typically smaller than the cloud radius, see Eq.~\eqref{eq_Field_length}.}
    \label{fig_condensation}
\end{figure}

Here we derive, by performing a few approximations, the basic timescale for the condensation of the hot intercloud medium on the BLR cloud surface. The basic set-up of the condensation is illustrated in Fig.~\ref{fig_condensation}. The accreted mass via the condensation from the hot phase can be expressed as $M_{\rm acc}\simeq 4\pi R_{\rm c}^2 \lambda_{\rm F} \rho_{\rm h}$, from which the accretion rate follows as $\dot{M}_{\rm acc}=8 \pi R_{\rm c} \dot{R}_{\rm c} \lambda_{\rm F} \rho_{\rm h}$, where we assumed that the cloud radius changes due to the condensation and the Field length remains constant. By assuming that the accretion of the condensed matter onto the cloud surface is approximately spherical with the characteristic (thermal) sound speed in the hot medium of $c_{\rm h}$, we may also write $\dot{M}_{\rm acc}=4\pi R_{\rm c}^2 \rho_{\rm h} c_{\rm h}$. By setting both expressions for $\dot{M}_{\rm acc}$ equal to each other, we obtain 
\begin{align}
    \frac{\dot{R}_{\rm c}}{R_{\rm c}} &= \frac{c_{\rm h}}{2\lambda_{\rm F}} \,,\notag\\
    \int_{R_{\rm c}}^{e R_{\rm c}} \frac{\mathrm{d}R'_{\rm c}}{R'_{\rm c}} &= \int_0^{\tau_{\rm cond}} \frac{c_{\rm h}}{2\lambda_{\rm F}} \mathrm{d} t\,,\notag \\
    \tau_{\rm cond} &= \frac{2\lambda_{\rm F}}{c_{\rm h}}\,,\notag\\
    &=2\left(\frac{\kappa_{\rm h} \mu m_{\rm p}}{k_{\rm B}n_{\rm h}^2 \Lambda_{\rm h}}\right)^{\frac{1}{2}}\sim 8.1 \times 10^{-4}\,{\rm yr}\sim 7.1\,\text{hours}\,.
    \label{eq_cond_time}
\end{align}

Hence, the condensation process is exponential and the BLR cloud radius evolves with time as $R_{\rm c}=R_{\rm c,0} \exp{(t/\tau_{\rm cond})}$, with the characteristic condensation timescale $\tau_{\rm cond}$ that is orders of magnitude shorter than the evaporation timescale. Although the exponential condensation model derived here is rather simplistic in terms of geometry and stationarity, it shows that the condensation can swiftly restore the BLR cloud material despite the short KH instability and tidal timescales. The condensation timescale is also plotted for comparison in Fig.~\ref{fig_tau_vel} alongside the flight time and other hydrodynamical timescales. 

\subsection{Objects Interacting with the Accretion Disk}

Let us note that shocks can also be produced in the accretion disk not only  by the passages of clouds but also stars present in a dense nuclear cluster \citep{1996Ap&SS.237..187D}. The relative velocity of these bodies with respect to the accretion disk medium can span a large interval. In the case of clouds emerging from the accretion disk, the majority of them is only weakly accelerated above the mid-plane, where they hover and eventually land back onto the disk surface at a similar radius with just a moderate component of the vertical velocity. On the other hand, there is a population of clouds that execute a more extended trajectory above the disk towards a different landing radius, where the azimuthal velocity component of the cloud is very different from the local orbital velocity of the underlying material.

In rare situations, the cloud motion can be influenced by the large-scale magnetic field twisted near the AGN jet axis. The cloud can then hit the disk surface in counter-rotation, which would further enhance the shock propagating from the impact place. The pressure-confined clouds thus make a significant difference from relatively weak tails behind stars passing across the accretion disk in the scenario described by \citet{zurek1994} and \citet{2004CQGra..21R...1K}. As further pointed out, turbulence in the star wake can significantly enhance the amount of matter that is pulled out from the accretion disk plane and contribute to the supply for the BLR region and also the enhanced growth rate of the central black hole \citep{1996ApJ...470..652Z,2005ApJ...619...30M}.

Particle acceleration and the associated non-thermal emission due to stellar-wind bow shocks could be relevant for the closest galactic nucleus--the Galactic center \citep{2005ApJ...635L..45Q}. Fast-moving S stars orbit the supermassive black hole with velocities of a few thousand km s$^{-1}$ and depending on the properties of their stellar winds, the synchrotron flux could marginally be detectable in the radio and the infrared domains \citep{2016MNRAS.455L..21G,2016MNRAS.455.1257Z}. With the current instrument sensitivity in the near-infrared domain, the light curve of the S2 star is constant within the uncertainties, and any non-thermal contribution is beyond the detection level \citep{2020A&A...644A.105H}. This is consistent with the presence of the hot and diluted accretion flow in the Galactic center, which is in contrast to the optically thick and geometrically thin disks in AGN where the BLR clouds can originate. 

We assumed an overall smooth distribution of density in the disk or torus. In realistic set-ups, however, instabilities may lead to spatially concentrated variations {\em within} the disk plane. In places of significant dilution, the falling clouds can penetrate deeper into the medium, and in the extreme case, the clouds can even traverse the equatorial plane into the other hemisphere. Such penetrating collisions of BLR clouds are similar to star-disk collisions and they could lead to the ejection of more material above the disk plane that could then contribute to wide-angle nuclear outflows \citep[see e.g.,][]{2021ApJ...917...43S}.

\bibliographystyle{aasjournal}
\bibliography{fallback}

\end{document}